# Reviewer Assignment Problem:
# A Systematic Review of the Literature

**Meltem Aksoy**                                             yontarm@itu.edu.tr
**Seda Yanik**                                               sedayanik@itu.edu.tr
*Istanbul Technical University,*
*Industrial Engineering Department,*
*Maçka, Istanbul, Turkey*

**Mehmet Fatih Amasyali**                                    amasyali@yildiz.edu.tr
*Yildiz Technical University,*
*Computer Engineering Department,*
*Davutpaşa, Istanbul, Turkey*

## Abstract

Appropriate reviewer assignment significantly impacts the quality of proposal evaluation, as accurate and fair reviews are contingent on their assignment to relevant reviewers. The crucial task of assigning reviewers to submitted proposals is the starting point of the review process and is also known as the reviewer assignment problem (RAP). Due to the obvious restrictions of manual assignment, journal editors, conference organizers, and grant managers demand automatic reviewer assignment approaches. Many studies have proposed assignment solutions in response to the demand for automated procedures since 1992. The primary objective of this survey paper is to provide scholars and practitioners with a comprehensive overview of available research on the RAP. To achieve this goal, this article presents an in-depth systematic review of 103 publications in the field of reviewer assignment published in the past three decades and available in the Web of Science, Scopus, ScienceDirect, Google Scholar, and Semantic Scholar databases. This review paper classified and discussed the RAP approaches into two broad categories and numerous subcategories based on their underlying techniques. Furthermore, potential future research directions for each category are presented. This survey shows that the research on the RAP is becoming more significant and that more effort is required to develop new approaches and a framework.

## 1. Introduction

The identification and assignment of experts is an emerging research subject that many researchers have intensively investigated in recent years. The goal of expert recommendation is to make it easier to find the correct person to ask a specific question and have them answer it for us. The task here may be to identify or assign experts with particular expertise in a topic. This paper focuses on the reviewer assignment, which is a unique area within the expert assignment.

Peer review is a regular professional procedure used to evaluate the quality and feasibility of scientific research papers and projects. The assignment of expert reviewers to project proposals or research papers, known as the Reviewer Assignment Problem (RAP), is a critical and challenging task in the peer review process of academic conferences, journals, and grant organizations.

In order to maximize the quality of reviews and ensure a fair and impartial evaluation, the following important points should be considered when assigning reviewers: (1) reviewers must be assigned to proposals based on their areas of expertise; (2) proposals must be evaluated by the most qualified reviewers in their respective subject areas; (3) reviewers should be eager and enthusiastic about examining proposals; (4) no reviewer should examine more than the agreed-upon number of proposals; (5) each reviewer should analyze the same number of proposals; (6) each proposal should be reviewed by a certain number of reviewers; and (7) relationships between applicants and reviewers that may undermine the fairness of the review process should be avoided.

Traditionally, reviewer assignments were conducted by a single decision-maker (such as a grant program manager) or a small committee, and all these tasks were performed manually. Nevertheless, the manual assignment process was highly time-consuming and entailed the subjective or biased judgment of the committee. In addition, it was challenging to optimize the assignments because all the requirements could not be taken into account properly. Moreover, reviewer assignments had to be made under severe time limitations as many applications were received close to the deadline that had been announced. In the current peer review process, there are many reviewers and proposals, even thousands. Such a high number of reviewers and proposals slows down the manual assignment, making peer review problematic for conference chairs, journal editors, and grant managers. Thus, the manual approach was extremely laborious and did not always produce the optimal outcome. Therefore, the approach of automatically assigning reviewers has become an essential topic in the academic community. The earliest study that addressed the RAP was proposed by Dumais and Nielsen (1992), which presented the similarity score calculation based on Latent Semantic Indexing (LSI). In response to the demand for automatic procedures, many studies have proposed assignment solutions since Dumais and Nielsen's (1992) breakthrough.

Several review articles on the RAP approaches have been published, but they do not reflect the most recent research advancements on this topic. Most of these studies focused only on one particular aspect of the RAP and did not address the problem comprehensively enough to provide the overall research perspective. For instance, Goldsmith and Sloan (2007) presented a survey in which they discussed several optimization criteria and some of the techniques employed by program chairs and current conference management tools.

In another survey, Kalmukov and Rachev (2010) investigated the available approaches for defining the competency of papers and reviewers. Kolasa and Krol (2011) investigated heuristic algorithms for the paper reviewer assignment problem. The first literature review focusing on all aspects of the RAP was presented by Wang et al. (2008) and cited 19 studies. An updated and more comprehensive version of this study was introduced in 2010 by Wang et al. This study cites 36 studies proposed as solutions to the RAP. Since these surveys were published more than a decade ago, they do not reflect the latest scientific developments in this field. Recent research by Patil and Mahalle (2020) addresses the generally used performance indicators for the RAP. Also, in all the surveys that have been done so far, only the problem of conference reviewer assignment has been dealt with. Studies of how reviewers are assigned in journals and grant organizations have not been taken into account.

Furthermore, there are also survey papers that focus on the whole review process and handle the RAP as a component of this process. Even though there is a section in these studies about how reviewers are chosen for proposals, the surveys are insufficient to offer the research perspective on the RAP as a whole. For example, Price and Flash (2017) provided an overview article of how machine learning and AI tools are used to automate steps of the peer review process. More recently, Shah (2022) presented a comprehensive survey article in which the challenges in the peer review process and the proposed solutions were investigated. This study provides a general overview of the literature on peer review by addressing the inequalities in the review process that result from issues such as dishonest behavior, miscalibration, and subjectivity across the entire assessment process. In this study, the RAP is handled in terms of injustice caused by incompatible reviewer expertise and has been examined superficially with only one dimension.

In order to address the research gap caused by the shortcomings of previous surveys, this study aims to establish an updated baseline of the RAP-related studies and present a comprehensive systematic literature review that covers all aspects of the RAP. It presents an in-depth review of 103 studies published between 1992 and 2022 and available on the Web of Science, Scopus, ScienceDirect, Google Scholar, and Semantic Scholar databases.

The primary contribution of this review is to provide readers with a thorough understanding of the RAP and current state-of-the-art approaches. In addition, this literature review proposes a basis for classifying available approaches addressing the RAP. Researchers can use this publication to consider the gaps in the literature and identify prospective study subjects for the future. Furthermore, grant organizations, scientific journals, and conferences can better understand how the reviewer assignment process can be managed more efficiently.

This research is organized as follows: The methodology used to conduct the systematic literature review is presented in Section 2. Section 3 presents the different application areas of the RAP. Section 4 gives stages of the reviewer assignment process. Sections 5, 6, and 7 examine and summarize the publications through the stages of the RAP. Section 8 concludes with recommendations for future research.

## 2. Review Methodology

The systematic literature review presented in this paper follows the guideline recommended by Snyder (2019), which consists of three main stages: (i) designing the review, (ii) conducting the review, and (iii) reporting the review. Figure 1 shows the plan for conducting this comprehensive literature review.

During the design phase, the research question has been determined, and an overall review approach has been considered, including selecting search terms and appropriate databases. The main research questions to be addressed include the following: what is the RAP, and what are the fields of study? What kind of information is used for modeling the reviewer profile and proposal profile, respectively? What are the different tools and techniques presented to solve the RAP? How to classify various tools and methods proposed to solve the RAP?

The search was conducted in five electronic databases: Web of Science, Scopus, ScienceDirect, Google Scholar, and Semantic Scholar. These databases were chosen because they provided the highest-impact publications and conference proceedings for the RAP. As search terms, "reviewer assignment" and "reviewer assignment problem" were used. This survey considered research articles (from journals and conferences) in English, published between 1992 and 2022, which could be found in digital databases. To ensure that the publications chosen were relevant to this survey, inclusion and exclusion criteria were identified to analyze each potential primary study. The criteria for this review are shown in Figure 1. To be considered eligible, a study must satisfy all inclusion criteria and none of the exclusion criteria.

During the review stage, a search strategy was used to overview the RAP literature comprehensively. The search strategy steps are as follows: 1. Review the titles, keywords, and abstracts of publications and choose the ones that meet the criteria for inclusion. 2. Read any publications that were not eliminated in the previous steps to determine if they should be removed from the survey based on the exclusion criteria. 3. Scan the bibliographies of eligible papers to discover new studies. If these studies are found, keep in mind that they must meet the inclusion and exclusion criteria.

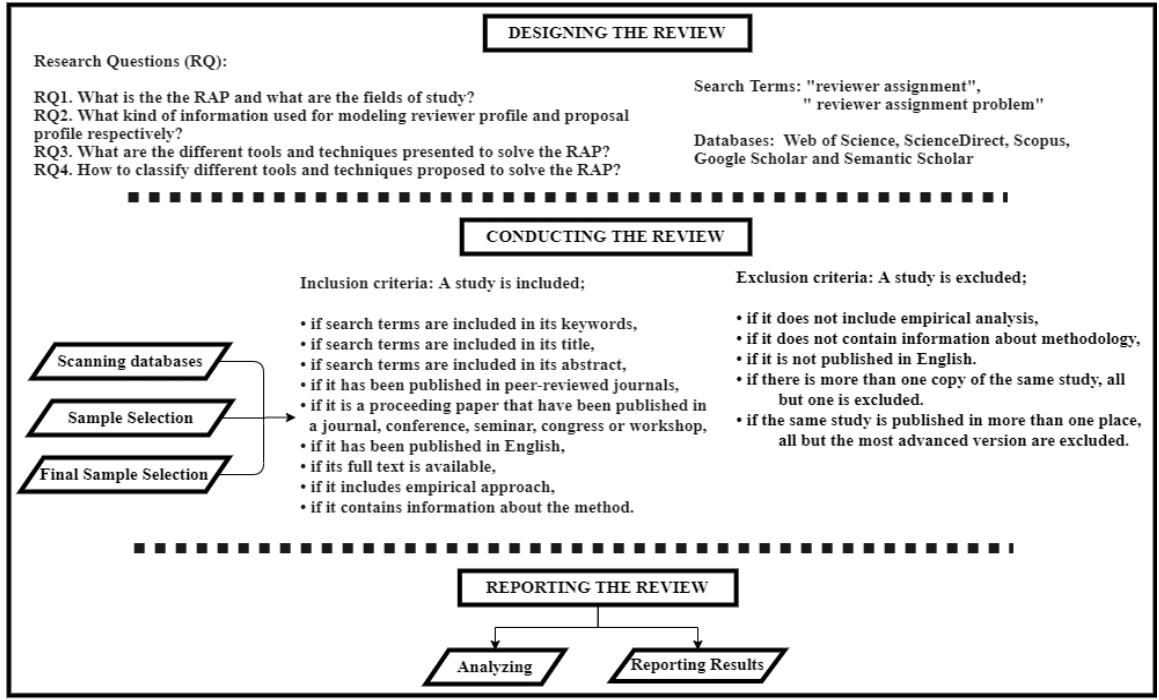

Figure 1: Plan of the systematic literature review.

In August 2022, the databases were searched, and the results were updated regularly. A total of 404 papers were identified by searching with determined terms. After the removal of duplicate studies using Mendeley, 334 papers remained. After identifying potential primary studies that addressed the RAP, the publications distribution by year and research fields were examined. As depicted in Figure 2, the overall distribution shows that the number of studies dealing with the RAP has increased in recent years.

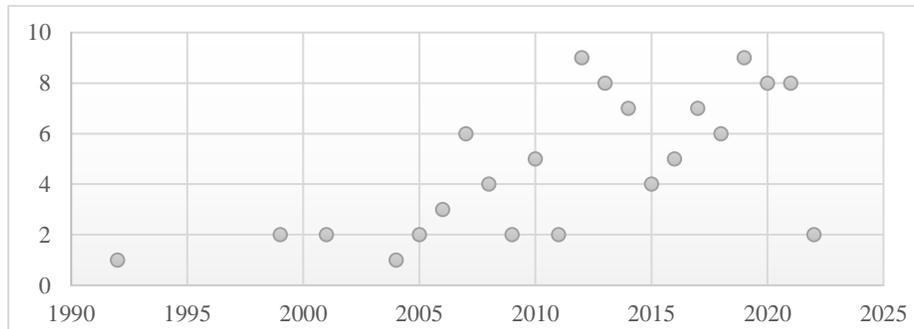

Figure 2: Distribution of studies per year.

The Web of Science categories was utilized to classify publications according to their research fields. Table 1 presents the distribution of studies per research field. Considering the nearly equal distribution of studies over a wide range of research fields, it can be concluded that the RAP is an important topic in several research domains and that the methodologies employed to address the RAP are diverse and multidisciplinary. As shown in Table 1, computer science information systems, educational research, and artificial intelligence are the first three research fields.

| Categories | Record Count | % of 334 |
|---|---|---|
| Computer Science Information Systems | 44 | 13,17 |
| Education Educational Research | 40 | 11,98 |
| Computer Science Artificial Intelligence | 36 | 10,78 |
| Engineering Electrical Electronic | 24 | 7,19 |
| Computer Science Theory Methods | 22 | 6,59 |
| Computer Science Interdisciplinary Applications | 20 | 5,99 |
| Education Scientific Disciplines | 19 | 5,69 |
| Information Science Library Science | 13 | 3,89 |
| Surgery | 13 | 3,89 |
| Computer Science Software Engineering | 12 | 3,59 |
| Operations Research Management Science | 11 | 3,29 |
| Pediatrics | 11 | 3,29 |
| Health Care Sciences Services | 10 | 2,99 |
| Medicine General Internal | 10 | 2,99 |
| Clinical Neurology | 9 | 2,69 |
| Telecommunications | 9 | 2,69 |
| Engineering Multidisciplinary | 7 | 2,10 |
| Public Environmental Occupational Health | 6 | 1,80 |
| Radiology Nuclear Medicine Medical Imaging | 6 | 1,80 |
| Biology | 5 | 1,50 |
| Computer Science Cybernetics | 4 | 1,20 |
| Orthopedics | 3 | 0,90 |

Table 1: Distribution of studies per research field.

The inclusion and exclusion criteria were then applied to each article's titles, keywords, and abstracts. In this step, five papers written in languages other than English were excluded, and three articles were excluded because full access was denied. As a result of this procedure, 74 studies were identified as worth reading in full text. After a deeper analysis, 65 studies met the criteria and were found eligible.

Through the reference and citation search of primary studies, 1768 articles were identified and examined according to exclusion and inclusion criteria. Subsequently, 60 studies were discovered to be potentially relevant to the subject of this survey. The full texts of 60 studies were examined deeply, and 38 of them were found eligible. A total of 103 papers satisfied the criteria and were considered for systematic review. Figure 3 depicts a synthesis of this procedure.

## 3. Application Areas of the RAP

Researchers responded to the requirement to automate reviewer assignment tasks with several approaches. To evaluate the performance of the presented approaches, they used different datasets from various application areas. Application areas of studies analyzed for this survey are listed in the second column of Appendices B and C. Furthermore, the distribution of these studies in terms of proposal type is shown in Figure 4. It is seen that most of the studies address the problem of assigning appropriate reviewers to conference papers (54%). The second most studied application area is the assignment of reviewers to project proposals (25%), and the third is the assignment of reviewers to journal papers (21%).

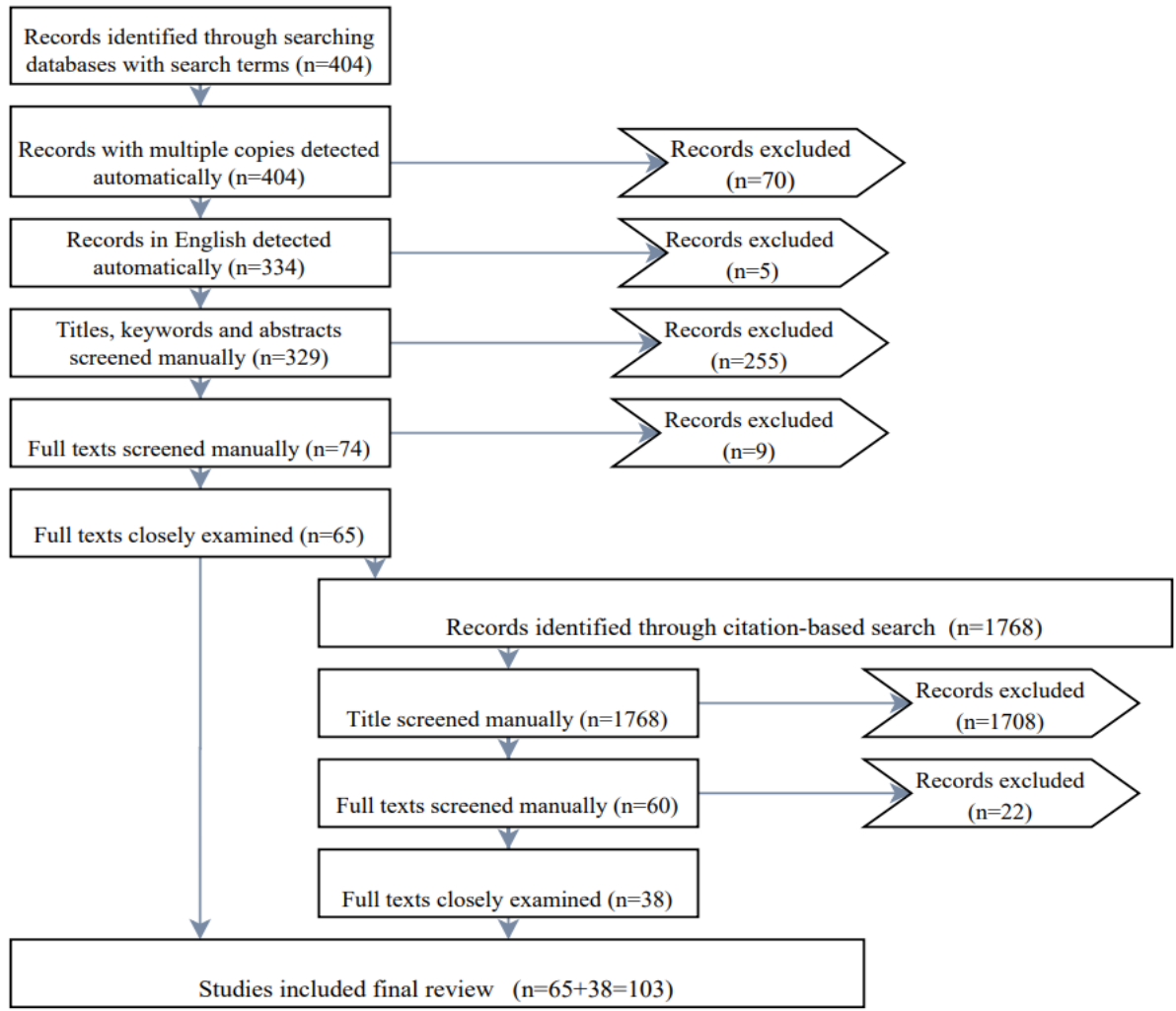

Figure 3: Study selection process.

Even though there are many empirical studies on the RAP in different application areas, it is difficult to find publicly available information about real reviewer assignments. The main reason is that academic journals, conferences, and grant organizations do not share reviewer and proposal information due to privacy concerns. Hence, some scholars created various datasets that have been used to assess the performance of proposed models and as benchmark datasets in various research. Accessible datasets are listed in Appendix A.

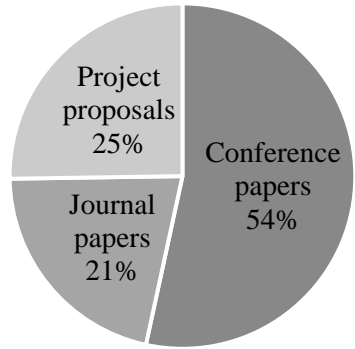

Figure 4: Distribution of studies per proposal type.

## 3.1 RAP For Conference Papers

The peer review process, widely used by many conferences, acts as a filter that separates eligible papers from those that are not and selects high-quality conference papers. The review procedure begins with the submission of the paper before a predetermined deadline. During the submission phase, various data on papers are collected to create a paper profile. After the submission deadline, the conference program chair must select expert reviewers to evaluate the submitted papers from the reviewer pool. Reviewers are frequently asked to "bid" on which papers they are willing to read. Appropriate reviewers are assigned to papers based on bidding rates and information provided by reviewers during the registration process, such as expertise area, affiliation, and seniority. Reviews assess the papers according to the review form of the conference and return review comments to the program chair. Generally, the program chair makes the final decision regarding whether or not to accept each paper based on the reviewer's comments. All authors often receive anonymous excerpts of the reviewer's comments so they can improve their papers. Finally, authors of papers that have been accepted may submit a revised version of their work in what is known as "camera-ready format." Even a small mistake in the assignment may result in serious misjudgments that can significantly diminish the conference's reputation and the author's confidence in the event. Therefore, selecting reviewers is the most crucial phase because it directly affects the quality of the entire review process (Shah, 2022).

During the conference peer review process, reviewers must be assigned to papers simultaneously, and the task of assigning reviewers must be done within a certain amount of time. Small conferences with few papers and reviewers can manually assign papers to reviewers. However, the number of submitted papers and reviewers is quite large at most conferences, and the papers are typically submitted close to the conference deadline. In these circumstances, manually assigning the best reviewer for each paper is challenging. In addition, manual assignment is becoming less accurate for large conferences due to many constraints (accuracy, load balancing, conflicts of interest, etc.) that need to be considered.

Automation has been in high demand in the review process for conferences to overcome the difficulties of the manual process. Automated reviewer assignment not only has the potential to facilitate the task but also leads to less biased and more systematic paper-reviewer matches than manual protocols. The request to automatically assign reviewers to papers for academic conferences has garnered the interest of numerous scholars. They have proposed many automatic reviewer assignment approaches, some of which have been implemented in actual conferences. The most well-known attempt toward a real-world implementation for conferences is the Toronto Paper Matching System (TPMS) (Charlin & Zemel, 2013). Since 2010, it has been utilized at conferences on machine learning and computer vision, such as NIPS, ICCV, and ICML. In addition, it was incorporated into the Microsoft Research Conference Management Toolkit, a web-based system for managing manuscript submissions and review procedures. Other well-known systems such as GRAPE (Di Mauro & Ferilli, 2005) and Erie (Li & Hou, 2016) are also used in the review assignment process of large conferences such as NIPS, ICML, and IEEE INFOCOM.

## 3.2 RAP for Journal Papers

The proper and accurate review of papers plays a crucial role in ensuring the quality of scientific journals. As is the case with conference papers, scientific articles should be evaluated based on specific criteria and allocated to reviewers so that they can be published following an objective evaluation. Regarding paper submission and evaluation, conference and journal management have many common points. However, the critical distinction between them is in the assignment of reviewers to publications. The journal management system should identify the most appropriate reviewers for a given paper at each time without treating the assignment as a global optimization problem like the conference management system. Moreover, conferences must do a one-time assignment, whereas journals must do a sequential one. There is a paper submission deadline for conferences. In contrast, journals often do not have deadlines, and papers are handled independently of each other. Although there is no deadline for journal submissions, reviewers are expected to complete their evaluations at a certain time.

For journal articles, editors and associate editors assign reviewers manually on a routine basis. Manual assignment strongly depends on editors' understanding of the most recent academic advancements on the subject. Due to the increasing number of submissions, finding the right paper-reviewer pairs under constraints has become a challenging and time-consuming task for journal editors. The system needs to be automatic for more fair, objective, and expedient review assignments. Despite this requirement, the number of empirical studies submitted to the RAP for journal papers (Biswas & Hasan, 2007; Karimzadehgan & Zhai, 2009; Daud et al., 2010; Andrade-Navarro et al., 2012; Wang et al., 2013; Protasiewicz, 2014; Yin et al., 2016; Peng et al., 2017; Jin et al., 2018a, 2018b; Zhao et al., 2018; Duan et al., 2019; Chughtai et al., 2019; Tan et al., 2021; Hoang et al., 2021) is less than that for conference papers.

## 3.3 RAP for Project Proposals

Because of the intense competition and limited funding sources, selecting appropriate projects to provide financial support is vital for grant organizations (Tian et al., 2002). Project selection is a complex and time-consuming procedure with numerous sub-tasks. One of the important and challenging tasks in this complex process is assigning reviewers to project proposals. Because their opinions impact the potential value of a project, reviewers play an essential role in project selection. The method used to assign the most qualified experts to examine project proposals may have a major impact on the quality of project selection and the grant organizations' return on investment.

RAP for project proposal evaluation has attracted the attention of researchers. Though, it is seen that the number of studies involving empirical analysis with real data is insufficient. National Science Foundation (NSF) (Janak et al., 2005; Hettich & Pazzani) and the National Natural Science Foundation of China (NSFC) (Xu et al., 2010; Mai et al., 2012; Li & Watanabe, 2013; Silva et al., 2014; Liu et al., 2016; Yue et al., 2017) stand out as two institutions whose datasets are used in many empirical analyses.

For the remainder of this study, the term proposal will refer to both paper and project proposals.

## 4. Stages of Reviewer Assignment Process and Related Methodological Approaches

Regardless of how the reviewer assignment is structured (manual or automatic), Wang et al. (2010) divided the procedure into three steps: (1) search for reviewer candidates, (2) calculate the similarity score between each proposal and each reviewer candidate and (3) optimize the assignment to maximize similarity score under constraints. Yin et al. (2016) regarded the reviewer assignment task as an expert retrieval problem, and the thematic relevance between reviewer candidates and proposals is the primary focus. From this perspective, they characterized the RAP process as follows: Firstly, information on potential reviewer candidates is gathered to represent one's expertise and knowledge. Secondly, proposals are modeled as queries. Finally, reviewers are chosen based on the relationship between their expertise and the contents of the proposals.

Two process definitions of the RAP, one defined by Wang et al. (2010) and the other by Yin et al. (2016), are combined in this systematic review study, and the RAP process is divided into three stages. Creating reviewer and proposal profiles using different information sources to represent candidate reviewers and proposals is regarded as stage (1), which is called the representation of reviewers and proposals in this survey. The primary goal of the stage (2) is to calculate a similarity score for every possible proposal-reviewer pair based on the profile of candidate reviewers and proposals. The similarity score between any proposal and reviewer is a number that captures the expertise match between the reviewer and proposal (Shah, 2022). A higher similarity score indicates a higher expected review quality. Given the calculated similarity score, the optimal assignment can be established in stage (3), considering practical constraints such as workload balance, proposal demand, and reviewer preferences.

To create the profile of the reviewer and proposal, all researchers analyzed the various information that might represent a reviewer or proposal before calculating the relevance between proposals and reviewers or optimizing it. That means all the studies reviewed applied the first stage. Nevertheless, it may not be the case for stages (2) and (3). There are many approaches presented to solve the RAP. While some studies focus only on calculating the similarity score between the reviewer and the proposal (e.g., Rodriguez & Bollen, 2008; Daud et al., 2010; Xu & Du, 2013; Protasiewicz, 2014; Mittal et al., 2020, etc.), others try to solve the assignment problem by optimizing objectives under some constraints (Cook et al., 2005; Li et al., 2008; Tang et al., 2010; Tang et al., 2012; Daş & Gökçen, 2014; Guo et al., 2018; Boehmer et al., 2021, etc.). Also, there are studies dealing with all stages of the RAP and designing it as a reviewer recommendation system or decision support system (e.g., Tian et al., 2005; Janak et al., 2006; Sun et al., 2007; Conry et al., 2009; Xu et al., 2010; Silva et al., 2014; Liu et al., 2016; Hoang et al., 2021).

As many researchers agree (Wang et al., 2008; Xu et al., 2010; Kou et al., 2015a; Peng et al., 2017; Duan et al., 2019; Yang et al., 2020), proposed approaches for stages 2 and 3 are divided into two main categories: i) the information retrieval-based reviewer assignment problem (IRRAP) and (ii) the optimization-based reviewer assignment problem (ORAP). The distribution of 103 publications into two main categories is presented in Figure 5. As depicted in Figure 5, some studies focus on only one approach, while others focus on both approaches. In this systematic review study, besides the representation of the reviewer and proposal, the studies were examined under two headings according to the proposed approaches to solve the RAP. Studies addressing both categories are discussed separately under each category according to the methods used.

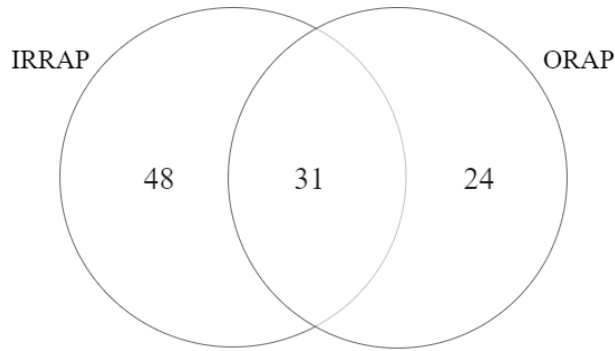

Figure 5: Distribution of studies per two main categories.

IRRAP focuses on the second stage of the RAP, with the primary goal of calculating the proposal-reviewer similarity score (Dumais & Nielsen, 1992; Mimno & Mccallum, 2007). The RAP is treated as an information retrieval problem, and the information retrieval-based methods consider each proposal as a query and a reviewer as a document to retrieve the appropriate reviewers. In this survey, IRRAP is divided into three groups based on how different types of information are used: text-based/content-based approaches, network-based approaches, and manual information-based approaches. The manual information-based approach assigns reviewers to the proposal with additional information precisely provided by reviewers, for instance, collecting reviewers' interests or ratings about proposals, allowing reviewers to provide topics about themselves, etc. This approach usually requires collecting manual information. Text-based/Content-based approaches generally use Natural Language Processing (NLP) techniques to statistically learn word co-occurrence information in documents, such as between words and topics (topic model), between words and documents (language model), and between words and words (word embedding models). Text-based/content-based approaches can also be divided into two categories of related methods: semantic-based and word-based. The network-based approach typically benefits from network relationships such as citation networks, co-authorship networks, collaboration networks, and other related networks between reviewer and reviewer or reviewer and proposal.

ORAP focuses on the third stage of the RAP, turns the IRRAP into an optimization problem, and tries to solve this problem from the perspective of operations research (Benferhat & Lang, 2001; Cook et al., 2005; Xu et al., 2010; Kolasa & Krol, 2011). In this study, ORAP was examined regarding the mathematical models and the algorithms used to solve the models.

All of the studies examined in this survey are summed up under three main headings based on the stages of the RAP process and the ways that the RAP can be solved. The representations of the reviewer and proposal are discussed in Section 5. Section 6 discusses information retrieval methods for calculating similarity scores between proposals and reviewers, while Section 7 discusses mathematical models and algorithms for optimizing reviewer assignments. The overall classification of the RAP approaches is presented in Figure 6. As seen in Figure 6, each class is labeled by a heading number. This paper surveys related studies in each title.

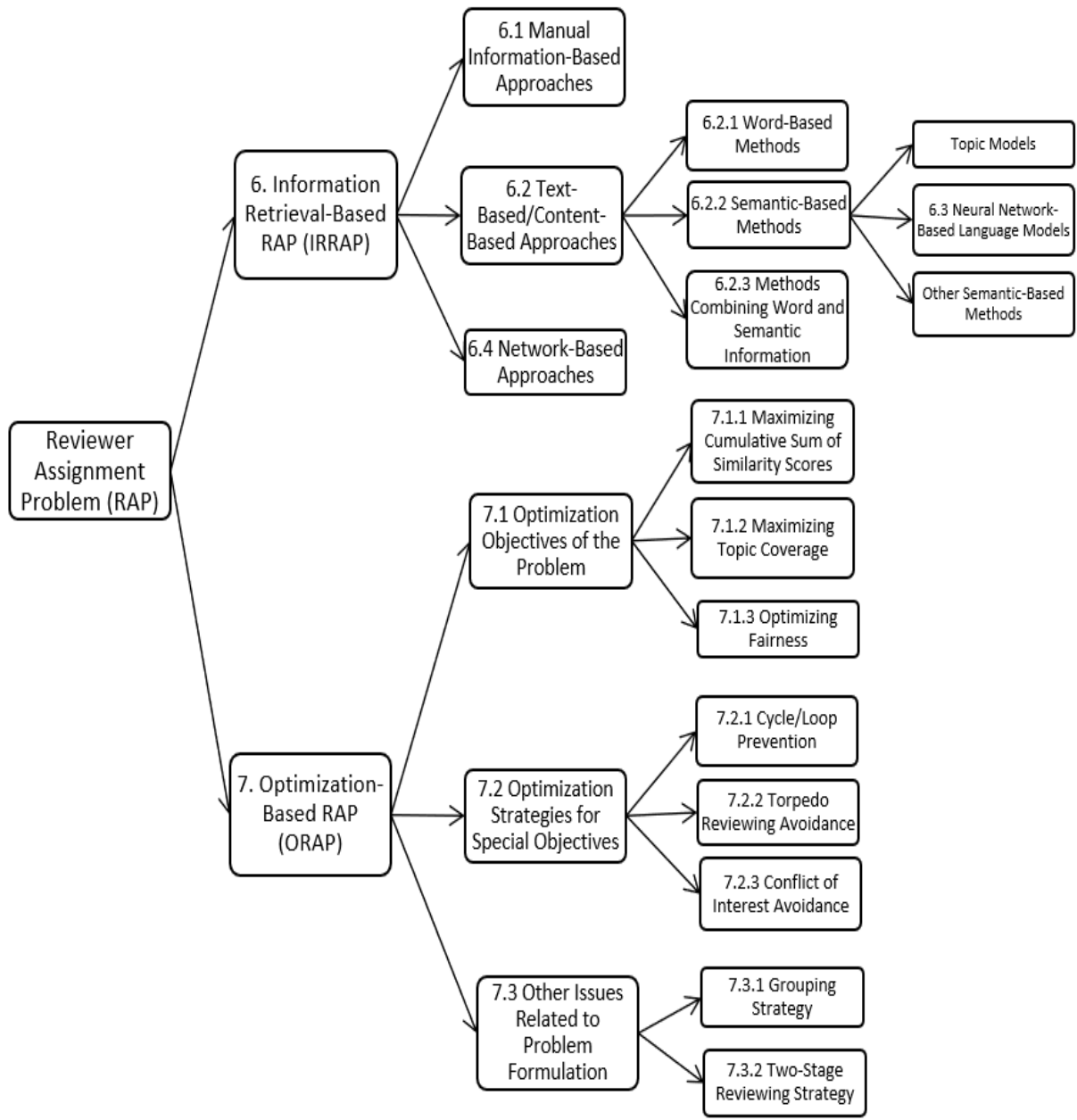

Figure 6: Classification of the RAP approaches.

## 5. Stage One: Representation of Reviewers and Proposals

Before computing the similarity scores between a proposal and a reviewer, two things are usually defined first: the reviewer's profile and the proposal's profile. Researchers look at several pieces of information that can be used to represent a reviewer and proposal. It is common to use information from the reviewer's published papers to create the reviewer profile, such as the title, abstract, keywords, full text, reference list, citation list, and metadata of past publications. In addition to this information, the profiles can include any personal information about the reviewers, such as their name and surname, affiliation, title, interest area, expertise area, h-index, citation network, co-author information, other relationship/connection information (e.g., academic relationships, colleagues, competitor relationships, advisee-advisor relationships, project collaborators, cooperations, etc.), reputation/seniority, past evaluation experience, proposal application experience, project management experience (number of projects, quality of projects), etc.

Also, self-declared keywords and reviewer preferences about proposals (self-evaluations of each reviewer for each proposal, the degree of willingness each reviewer gives for each proposal, and interest in the proposal) can be used to represent reviewers.

The distribution of the publications per the information used to create the reviewer's profile is presented in Figures 7 and 8. According to Figure 7, the three most commonly used personal information to profile the reviewers are expertise/field of research, co-authorship information, and affiliation. The main reason for choosing this information is that it is easily accessible. This type of information is usually provided by reviewer candidates when they apply to be reviewers. But the reviewers may not be fully represented by their expertise/area of research, co-authorship information, or affiliation. In addition, reviewer candidates are frequently unable to appropriately choose their expertise area or research field due to their subjectivity and the possibility of misinterpretation. For this reason, benefiting from the rich information of reviewers' published papers provides much more accurate results.

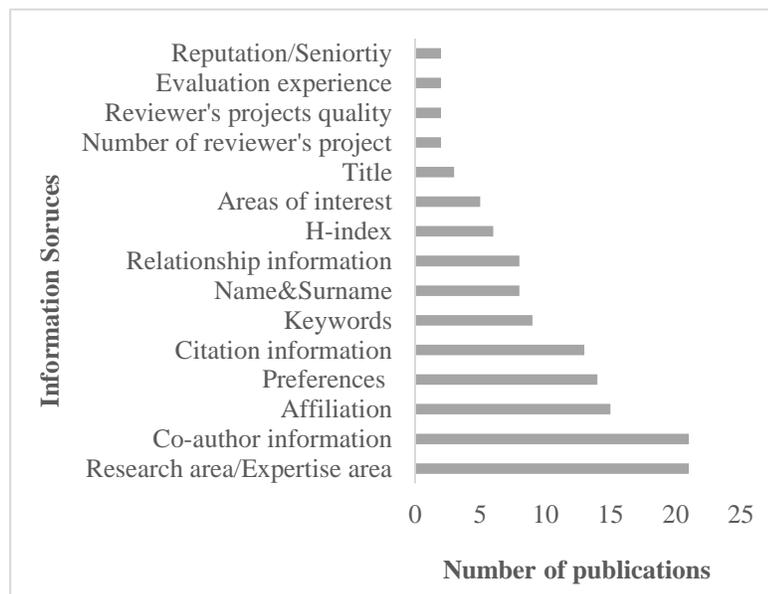

Figure 7: Number of publications per information source related to reviewers.

As seen in Figure 8, the three most preferred information sources about reviewers' published articles are full text, abstract, and title. Although only a single piece of information can be used to build a reviewer's profile, it is quite common to use multiple pieces of information to represent reviewers accurately. For example, the abstract and title of a reviewer's previous publications are frequently combined to form their profile.

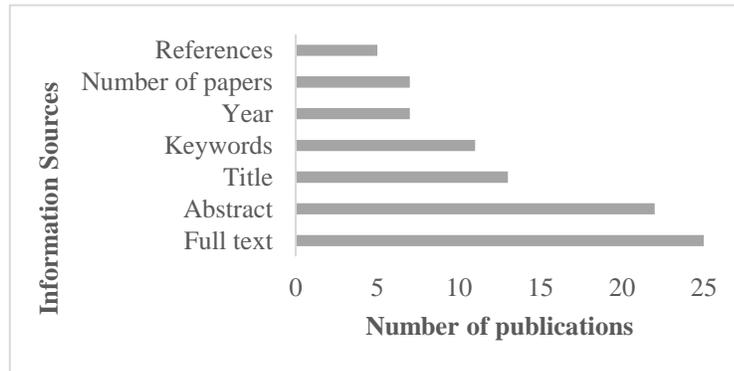

Figure 8: Number of publications per information source related to the reviewer's published papers.

When creating the proposal's profile, information related to the proposal, such as title, keywords, abstract, full text, subject, research/expertise area, reference, citation information, co-author information, as well as information such as applicant's institution, discipline/specialty, can be used. In addition, keywords specified by the applicant during the submitting phase can be used. The distribution of studies according to the information used to create the proposal's profile is presented in Figure 9. The three most commonly used information are the proposal's abstract, keyword, and full-text. Full-text usage was more common in studies after 2014 (73%). The main reason for this is thought to be the new generation of high-performance text/word representation techniques in NLP.

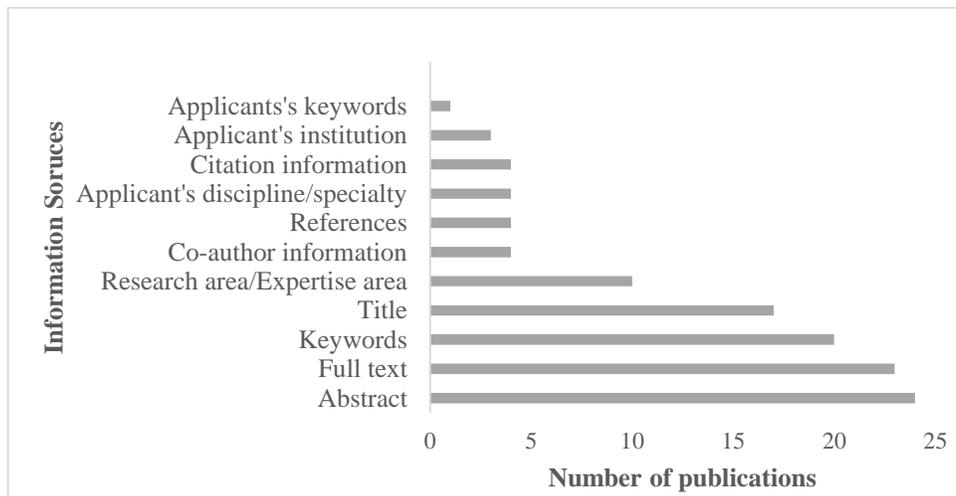

Figure 9: The number of publications per information source related to the proposal.

Detecting potential conflicts of interest (COIs) between the proposals and review is essential during proposal and reviewer profiling. Any cooperation or competition between the two parties could result in injustice. Consequently, potential COIs should be detected and avoided in the assignment task. Hence, studies on the detection of potential COIs between proposals and reviewers are discussed in a subsequent section.

## 5.1 Conflict of Interest Detection

Detecting COIs is crucial for ensuring an unbiased and fair proposal evaluation process. There are various kinds of COIs that should be detected and avoided in the reviewer assignment task. Wu et al. (2018) divided COIs into two distinct groups: definite COIs and latent COIs. Definite COIs are related to definite relations that proposals and reviewers can state in accordance with pre-defined declaration rules. Examples of definite COIs include co-authorship, colleague relationships, advisee-advisor relationships, project collaborators, and competitor relationships. Latent COIs, on the other hand, relate to relationships that are not needed to be stated according to the declaration rules. They must be inferred from explicit relationships, such as close friends, academic collaborators, family relatives, etc. In current well-known conference management systems, such as EasyChair and Microsoft Conference Management Toolkit, it is common to use definite COIs. These systems typically ask reviewers and applicants to disclose any COIs by looking at a long list of potential reviewers. In some cases, this list may contain several hundred reviewers. Also, authors may intentionally or unintentionally overlook some potential cases of COIs. So, the self-declaration procedure is time-consuming and potentially incomplete. Even though program committee chairs (at conferences), editors (at journals), or grant managers (at grant organizations) can check for potential COIs on their own, this is a time-consuming, insufficient, and error-prone process. Therefore, there is a need for effective and efficient methods to detect COIs automatically.

In some cases, it is difficult to detect COIs due to a lack of necessary information. Nevertheless, in many other cases, implicit or explicit information exists in the form of social networks, such as research social networks (Scholarmate), co-author networks (e.g., ArnetMiner, DBLP), and professional networks (e.g., Linkedin). To reduce potential COIs among the proposals and reviewers as well as those among the reviewers, Yin et al. (2016) integrated the probabilistic topic model and activation spread model into the researcher's co-authorship network. Also, Pradhan et al. (2020) used the co-authorship graph for COIs detection. They extracted a co-author list for each reviewer from ArnetMiner. They cross-checked it with the reviewer's and author's self-declared co-author relationship. Silva et al. (2014) created the collaboration network by mining ScholarMate data to identify potential COIs between reviewers. Yan et al. (2017) constructed academic networks of researchers and their institutions using academic activity data. Relationships between researchers include colleague, co-author, and advisee-advisor relationships, whereas institution-to-institution relationships consider cooperation and co-worker relations. They detected latent COIs using the path distance between proposals and reviewers in these academic networks.

Earlier research advocated leveraging weighted relationships to detect latent COIs automatically. Due to the lack of a general definition of COIs, it isn't easy to obtain the ground truth in COIs scenarios. Wu et al. (2018) presented a semi-automatic system called PISTIS (Platform for ConflIct of IntereST-aware Reviewer Suggestion) to address the COIs declaration and detection problem. PISTIS retrieves heterogeneous associations from various public sources, including co-authorship (from DBLP) and co-working duration (extracted from the affiliations). The advisor-advisee relationship is also extracted using a factor graph model (Wang et al., 2010). This heterogeneous information is used to construct a heterogeneous publication network that serves as the system's primary knowledge base. To simplify the declaration process, the system recommends latent COIs by a supervised ranking model that can be iteratively updated using historical declaration data.

## 6. Stage Two: Similarity Score Computation Between Proposals and Reviewers

The second stage of the RAP, the computation of the similarity score between proposals and reviewers, is typically defined as the Information Retrieval-based Reviewer Assignment Problem (IRRAP) in the literature. The information retrieval-based methods consider each proposal as a query to retrieve the appropriate reviewers. In this paper, the existing retrieval-based reviewer assignment studies were classified into three categories according to the usage of different types of information as follows: (i) manual information-based approach, (ii) text-based/content-based approach, and (iii) network-based approach. The distribution of the studies according to these three categories is presented in Figure 10. As depicted in Figure 10, some studies focus only on one category, and others address multiple categories simultaneously. Later in this part, the publications examined in this study are presented under three headings.

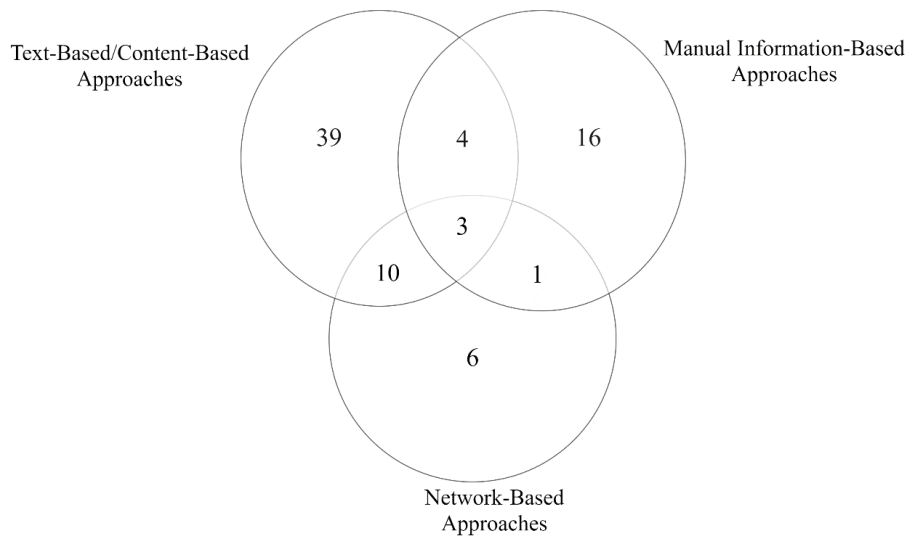

Figure 10: Distribution of IRRAP approaches per subcategories.

### 6.1 Manual Information-Based Approaches

The manual information-based approach assigns reviewers to proposals using additional information provided by reviewers. Bidding, also known as rating, is the most prevalent strategy in manual-based systems, in which the reviewer is involved in the process of allocating their preferences (Benferhat & Lang, 2001; Basu et al., 2001; Janak et al., 2005; Goldsmith & Sloan, 2007; Papagelis & Plexousakis, 2008; Xu et al., 2010; Garg et al., 2010; Kolasa & Krol, 2011; Chen et al., 2012; Daş & Gökçen, 2014; Tayal et al., 2014; Liu et al., 2015; Liu et al., 2016; Yeşilçimen & Yıldırım, 2019; Jecmen et al., 2020; Fiez et al., 2020). In the bidding method, reviewers must review the abstracts or the full text of the proposals and assign scores for each proposal representing their review willingness. Generally, the bidding scale consists of 3, 4 or 5 levels (Huang et al., 2010; Wang et al., 2010). For instance, in ICDM'07, the given bids range from 1 to 4, representing the preference as follows: 4 = "Eager to review", 3 = "Willing to review", 2 = "Not that willing to review", 1 = "Unwilling to review". After the bidding phase, each reviewer has a preference list for all the proposals, which represents to similarity score between the reviewer and the proposals. A higher bidding value indicates a higher similarity score. The expert is the best judge of whether or not he has sufficient competence in a proposal. In accordance with its intended use, it is seen that the bidding improves the quality and reliability of the evaluation and guarantees the reviewers' satisfaction when conducted by fair and responsible reviewers (Cabanac & Preuss, 2013). However, the bidding method has obvious disadvantages. First, when the number of proposals increases, the bidding becomes complex and takes too much time. Second, many conferences suffer from a lack of sufficient bids for the majority of applications. For example, 146 out of 264 proposals to the JCDL 2005 had no positive bids (Rodriquez & Bollen, 2007), and in IMC 2010, 68% of proposals received no bids (Beverly & Allman, 2012). If an insufficient number of proposals are bid on, the resulting proposal-reviewer bidding matrix will be sparse, which will drastically lower the accuracy of the reviewer assignment method. Third, reviewers tend to bid based on their current research interests rather than their cumulative expertise. This situation may reduce the quality of the assignment as well as the accuracy of the evaluation of the proposals. The last but most important disadvantage is that these approaches are open to malicious behaviors and conflicts of interest that will affect the fairness of the evaluation. With these approaches, reviewers can significantly increase their chances of being assigned to a proposal they can target by bidding strategically (Jecmen et al., 2020; Wu et al., 2021).

Expertise keywords or topics provided by reviewers and applicants are other manual information used in the assignment process (Hartvigsen et al., 1999; Xu et al., 2012; Charlin et al., 2012; Liu et al., 2014; Silva et al., 2014; Liu et al., 2015; Liu et al., 2016; Pradhan et al., 2020). The applicants are usually required to provide a short list of keywords for representing their proposals during submission. Similarly, during the registration process, reviewers are required to select the keywords corresponding to their expertise area. The keyword list can be created by selecting from a predetermined dictionary, such as the ACM Computational Classification System, or determined as free text. A set of keywords represents each proposal and each reviewer. One straightforward way of calculating the similarity score between two sets is by counting the number of keywords they have in common (Hartvigsen et al., 1999; Xu et al., 2012). Also, Jaccard similarity and Dice similarity measurements based on the common keywords can be used to compute similarity scores (Silva et al., 2014; Liu et al., 2015; Liu et al., 2016). Some researchers determined the assignment using both the preferences of reviewers and the keywords provided by reviewers (Di Mauro et al., 2005; Conry et al., 2009; Huang et al., 2010). But self-claimed keywords are subjective, just like bidding, and it could lead to misunderstandings and conflicts of interest.

In parallel with the disadvantages of manual information-based approaches, current studies that handled manual-based methods are less than those that aim to automate the RAP, such as text-based approaches. However, in some studies, for testing proposed techniques, manual-based information retrieval approaches are used as a kind of comparison (e.g., the comparison of proposal-reviewer matching based on bidding and based on the LSI (Dumais & Nielsen, 1992), comparison of bidding with the relatively ordered particle swarm algorithm (Rodriguez & Bollen, 2008), comparison of expert topics provided by reviewers and the information obtained by PLSA (Karimzadehgan & Zhai, 2012).

## 6.2 Text-Based/Content-Based Approaches

Text-based or content-based approaches are a kind of information retrieval technique that makes recommendations by comparing the properties of the content of a reviewer with those in which the proposal is interested (Davoodi et al., 2013). These methodologies are implicit methods used to identify proposals and the competencies of reviewers without requiring users to take any additional action utilized and to calculate the similarity score between reviewers and proposals via text mining technology.

In this section, text-based/content-based approaches are divided into two classes according to the different types of information used, inspired by Tan et al. (2021): the word-based methods and the semantic-based methods. Although studies that focus on semantic-based and word-based methods alone are in the majority, some studies handled both semantic-based and word-based methods simultaneously. Later in this section, studies dealt with word-based methods, semantic-based methods, and methods combining semantic and word information are presented under three headings. Besides, a detailed summary of these studies is presented in Appendix B.

### 6.2.1 WORD-BASED METHODS

Through statistical word frequency information, word-based methods are utilized to describe the relationship between the reviewer and the proposal. Word-based methods are presented in the following two categories as vector space models and statistical language models.

*Vector Space Model (VSM):* In the vector space model (VSM), each document of proposals and reviewers is represented as the vectors of a bag-of-words weighted by the TF/IDF score of words in that document. The word vector refers to a representation of each word as a numerical vector that corresponds to how the word is used or what it means. The similarity score between the reviewers and proposals is calculated according to vector representations of word information.

The VSM was first used by Yarowsky and Florian (1990), and they computed the similarity score between the reviewer and proposal by cosine similarity. Basu et al. (2001) proposed three different approaches based on VSM for the similarity score calculation: TF/IDF representation of the full text of documents, machine learning-based keyword extraction to represent both proposals and reviewers as keyword vectors, and domain ontology-based topic extraction to precisely define proposals and reviewers on a topic vector. The results of three different approaches showed that using more information leads to better performance. Hettich and Pazzani (2006) proposed a recommendation approach based on TF/IDF weighted vector space to recommend review panels for grant applications of the NSF. Caldera et al. (2014) introduced another approach that employs TF/IDF to categorize submitted articles in relation to existing publications of possible reviewers in order to automate the reviewer appropriateness rating. Automatic proposal-reviewer assignment using free-text but supplemented with keyword extraction and ontology-driven topic mapping offers better relevant proposal-reviewer matches in most circumstances. To prove this, Biswas and Hasan (2007) implemented TF/IDF to create a reviewer profile by utilizing both researchers' previous publications and field ontologies. Unlike other studies, Shon et al. (2016) did not directly use the TF/IDF value in the similarity score calculation between reviewers and proposals. Instead, the TF/IDF value obtained from the frequencies of the words in the reviewer's profile and the proposal's profile was used in the fuzzy weighted keywords.

Although researchers have demonstrated that VSM-based techniques can be used to assign submissions to reviewers without human interaction, the VSM has a fundamental disadvantage leading to an irrelevant proposal-reviewer match. VSM is a lexical matching technique based on the assumption that words are self-contained. This unrealistic assumption causes a loss of context and adjoining word integrity and ignores the grammatical structure of the sentence. The n-gram can be used to overcome the obstacles of the VSM. However, as n increases, the vector size increases exponentially due to the number of token combinations.

*Statistical Language Models (SLM):* Statistical Language Models (SLM) uses many probabilistic and statistical techniques to predict the probability distribution of linguistic elements in the sequence, such as words, sentences, and so on. In the RAP, SLM analyzes bodies of the text of proposals and reviewers and predicts the degree of similarity between them. Mimno and McCallum (2007) proposed the SLM with Dirichlet smoothing for measuring the affinity of a reviewer to a proposal. Mai et al. (2012) developed a new system based on a probabilistic language model to select suitable reviewers for each proposal. For the widely-used TPMS, Charlin and Zemel (2013) employed the SLM to compute the similarity score for each proposal-reviewer pair. Similarly, Tang et al. (2010) and Tang et al. (2012) used an SLM-based retrieval approach to calculate the pairwise similarity score. Karimzadehgan et al. (2008) introduced a mixed SLM to model multiple aspects of reviewer expertise.

Despite the practical usableness of SLM with the smoothing methods, data sparsity and the curse of dimensionality are compelling problems. Language modeling uses data-driven methodologies to capture the key statistical properties of natural language text sequences, which can be used to forecast future words in a sequence or conduct slot-filling in related tasks (Minaee et al., 2022). The link between consecutive tokens is captured by the simplest statistical language models of n-gram models. On the other hand, these models frequently fail to account for the long-distance reliance on tokens that encode semantic links.

### 6.2.2 SEMANTIC-BASED METHODS

The problem with current word-based methods is that users fail to access the information they are looking for without using the words in which the information is indexed. It is due to several complex features specific to the natural language. One of the most common is synonymy, which refers to a notion that can be expressed in a variety of ways (words). Another is polysemy, which means that many words have several meanings, allowing phrases in a user's query to match terms in documents that are not semantically relevant to the user. In short, although word-based methods can characterize the word-level frequency features of proposals, they cannot describe semantic information. Therefore, the similarity score calculated by these models may have a large deviation in some cases. To address these problems, semantic-based approaches are proposed. Semantic-based methods are presented in the following two categories, namely topic models and other semantic-based models.

### Topic Models

Studies based on semantic information are usually focused on topic models. Topic models are statistical models that are used to capture topic distributions in each reviewer's publications and submitted proposals, which aids in understanding the concentration of proposals and the expertise of the reviewer (Kim & Lee, 2018). Many topic modeling techniques, such as Latent Semantic Indexing (Dumais & Nielsen, 1992), Probabilistic Latent Semantic Analysis (Karimzadehgan et al., 2008; Mirzaei et al., 2019), Latent Dirichlet Allocation (Blei et al., 2003), Author Topic Model (Karimzadehgan & Zhai, 2009; Tang et al., 2012; Liu et al., 2014; Kou et al., 2015a, 2015b) and Author Person Topic Model (Mimno & Mccallum, 2007) are used to extract the defined topic distributions for reviewers and proposals.

*The Latent Semantic Indexing/Latent Semantic Analysis (LSI/LSA):* Latent semantic indexing (LSI) or latent semantic analysis (LSA) (Deerwester et al., 1990) is a commonly employed dimensionality reduction approach for assessing the similarity between two documents. LSI uncovers relationships between words and their underlying concepts based on a mathematical method named Singular Value Decomposition (SVD). The first study to use the LSI method to solve the RAP is proposed by Dumais and Nielsen (1992). They extracted the topic representation of reviewers and proposals as points in the k-dimensional LSI space. Then the similarity score between each proposal-reviewer pair was measured by the cosine distance of these two points. Ferilli et al. (2006) developed an expert system, GRAPE (Global Review Assignment Processing Engine), to automate the RAP. This system takes the proposals' topics and the reviewers' expertise and preference as input. Li and Hou (2016) created and implemented a review assignment system (Erie) for IEEE INFOCOM 2015 and INFOCOM 2016. In Erie, the matching scores for the proposal and reviewer candidates were computed using LSI. They benefited from the LSI method to extract topics for proposals and the expertise of reviewers. In the ontology-based recommendation system developed by Chughtai et al. (2019), ontologies were extracted and weighted using probabilistic approaches and compared on the basis of LSI for relevance index.

LSI is language-independent and uses a purely mathematical approach; thus, its computation method is relatively straightforward, efficient, and easy to use. However, it has a few major drawbacks (i) LSI needs a large corpus of documents and vocabulary to get accurate findings. Due to the high amount of data, additional storage and computation time are required. (ii) The topic information obtained using LSI lacks well-defined probabilities and is, therefore, hard to interpret. (iii) It is difficult the determine the optimal number of topics. (iv) SVD is computationally demanding and difficult to update as new data arrives. (v) LSI vectors take up far too much space.

*Probabilistic Latent Semantic Analysis (PLSA):* To overcome the disadvantages of LSI, Hofmann (1999) developed Probabilistic Latent Semantic Analysis (PLSA). PLSA, which may be stated as a probabilistic interpretation of LSA, attempts to determine which texts and words correspond to which topics based on the probabilities involved. PLSA offers a better interpretation of dimensions in latent spaces compared to LSI. Also, PLSA has been shown to provide a better model for capturing polysemy and synonymy than LSI (Farahat & Chen, 2006).

Although PLSA is preferable to LSA, it has not been utilized extensively for the RAP solutions. It was first implemented by Karimzadehgan et al. (2008) to extract different aspects of the reviewers' expertise. Then, the reviewers are paired with the proposals based on these aspects. In a recent study by Mirzaei et al. (2019), PLSA was used to obtain latent topics from the reviewers' previous publications and proposals. Due to several drawbacks of the model, PLSA can only be applied to a limited extent to the RAP. PLSA is not a well-defined topic model for the generation of new documents. Also, it uses the Expectation-Maximization (EM) algorithm for the model parameter training. The EM algorithm is driven by the likelihood function that is generated based on the data being observed and estimates the parameters by maximizing these likelihood functions by iteration. Consequently, the performance of the trained model is highly varied and largely dependent on the initial values. Another problem is that the number of parameters increases directly with the number of documents, resulting in an overfit model.

*Latent Dirichlet Allocation (LDA):* Latent Dirichlet Allocation (LDA) was created as a remedy for the primary deficiency of PLSA, which is that it lacks a well-defined generating model for new documents (Blei et al., 2003). LDA is a topic model that represents documents as a random mixture of latent topics identified by words. The Dirichlet distribution is utilized by LDA to determine the odds that texts and phrases belong to particular subjects. The model captures the interchangeability of both words and texts with the Dirichlet distribution, allowing for a cohesive test data generation procedure. It is seen that LDA is widely applied to extract topics representation from the reviewer's published studies and the submitted proposals (Mimno & Mccallum, 2007; Tang et al., 2010; Tang et al., 2012; Charlin & Zemel, 2013; Liu et al., 2014; Xu & Zuo, 2016; Peng et al., 2017; Yan et al., 2017; Nguyen et al., 2018; Medakene et al., 2019; Pradhan et al., 2020; Yang et al., 2020; Patil & Mahalle, 2021; Tan et al., 2021; Hoang et al., 2021).

As a state-of-the-art topic model, the LDA method successfully identifies the RAP research topic. Nevertheless, this technique has two fundamental disadvantages due to its assumptions. The first assumption is that the topic number must be fixed and known in advance. This limitation can be partially overcome with algorithms developed to determine the most suitable number of topics. The second assumption refers to the topical independence of the document. A document in the corpus may contain multiple topics, or the same topic may be present in multiple documents. However, in the LDA method, the topic ratios in the papers are random variables with a Dirichlet distribution and must be independent. Due to the assumption, LDA does not utilize the potential semantic relationship between words within a topic.

*Author-Topic Model (ATM):* Compared to other topic models, the traditional LDA has shown significant improvements in proposal topic extraction and reviewer and proposal-reviewer similarity score calculation. However, LDA does not account for author information. Considering the inadequacy of LDA for modeling the interests of authors, some baseline approaches were proposed as extensions of LDA. Author-Topic Model (ATM), developed by Rosen-Zvi et al. (2004), is the most widely used variation of LDA for proposal-reviewer matching.

Assuming that a document has many known authors and each author is regarded as a mixture of topics, the Author-Topic Model (ATM) attempts to determine which author is responsible for a particular word in the document. Kou et al. (2015a, 2015b) used the ATM to determine the topic distribution of proposals and reviewers' publications. Then, they provided expert group recommendations for each proposal according to the different topic weights. Taking into consideration the research interests of reviewers besides other commonly focused aspects such as the competence of reviewers and proposal-review similarity, Jin et al. (2017) used the ATM model and EM algorithm to estimate the topic distribution of proposals and publications of reviewers.

Several new LDA-based topic models for the RAP were introduced by incorporating different types of information into ATM. Tang et al. (2008) enhanced the ATM by including the publication venue and proposed the Author-Conference-Topic (ACT). While the meaning of a given topic remains virtually unchanged, the occurrence and correlates of topics (semantically related phrases) change over time. Based on this idea, Daud et al. (2010) proposed the Temporal-Expert-Topic (TET) model, which can identify experts on specific topics for different periods and show how the interests and relationships of experts have changed over time.

To determine the topic distance between reviewers and proposals, Mimno and Mccallum (2007) introduced the Author-Persona-Topic (APT) model. The APT model introduces a persona layer to ATM. Instead of collecting all of a particular author's papers into a single topic distribution, APT allowed each author's documents to be separated into one or more groups. These groups indicate various "personas" that a single author writes under, and each one is represented as independent topic distribution. Li and Watanabe (2013) extended the APT approach by adding a time component to the model.

In the ATM, the topic distribution of each author is considered only based on their documents, and the similarity of documents from different authors is ignored. Such limitations affect overall performance in modeling author interest, especially when analyzing documents in interdisciplinary studies. Jin et al. (2018a) proposed Author-Subject-Subject (AST) model to overcome the difficulties experienced in assigning experts to interdisciplinary proposals. AST model extends the ATM model by adding a subject layer. This way, the creation of hierarchical topics can be controlled, and the sharing of topics among authors is allowed.

Though it is claimed that topics in latent space are semantically coherent, topic models examine just the distribution of word cooccurrence in the document corpus and not the semantic information contained within them (Ekinci & Omurca, 2020). In addition, probabilistic topic models need a large document corpus to identify the topics and the distribution of topics precisely. This can be a problem when applied to short documents like abstracts (Anjum et al., 2019). Also, existing probabilistic topic models assume that words are independent and uncorrelated. While this assumption improves processing efficiency, it eliminates the rich relationships between words. Due to their limitations, semantically connected, meaningful and coherent topics cannot be generated using solely topic models.

## Other Semantic-Based Methods

Other semantic-based techniques are presented besides the topic models. Unlike the reviewer and proposal representation methods used in other studies, Li and Chen (2011) developed the similarity calculation between proposals and reviewers based on knowledge set theory. Müngen and Kaya (2018) believed that looking at the background information and metadata of reviewer candidates' research is insufficient to search for reviewers. Therefore, they introduced a new statistical approach called Topical Affinity Propagation (II-TAP), in which the reviewer's expertise was determined based on various features, such as the reviewer's papers, the metadata of these papers, the reviewer's changing interests over time and the reviewer's co-authors' expertise. In addition, they developed a strategy for eliminating the inadequacies in the abstracts and keywords of the papers used in the application. First, the papers' abstracts and keywords were obtained using the TextRank algorithm and the Zemberek-NLP3 library. Then, the frequency-finding module was used to complete the missing keywords.

Some researchers claim that it is not possible to properly determine the expertise level of reviewers in a particular field and the level of competence of reviewers varies by field and sub-field (Xue et al., 2012; Daş & Gökçen, 2014; Tayal et al., 2014; Mittal et al., 2019). Therefore, describing expertise levels as a crispy set may lead to inaccuracy in proposals-reviewers matching. Due to the difficulty of expressing the level of expertise in crisp values, in some studies, the similarity score is defined using linguistic variables to indicate the expertise of each expert on each proposal. Daş and Gökçen (2014) determined the similarity scores using linguistic variables to signify the expertise of each reviewer in relation to each proposal. In that study, linguistic variables are described by triangular fuzzy numbers. In another study, Tayal et al. (2014) used type-2 fuzzy sets to estimate the expertise level of the reviewers in the different fields. In addition, a fuzzy set of proposals was produced using the three terms that best represent proposals. With the help of fuzzy functions, the equality of the type-2 fuzzy set of reviewers and proposals is computed.

Sometimes, reviewers are unable to provide detailed information about their various areas of expertise manually. In turn, this may make it difficult to determine the whole region that encompasses all the relevant disciplines, decreasing the scope of assigning the correct reviewer to the proposal. Fuzzy logic extraction of keywords from the reviewer's past articles could be a feasible option. Andrade-Navarro et al. (2012) designed a web tool called Peer2ref to aid journal editors in identifying qualified reviewers. The tool extracts important concepts as keywords from the title and abstract of proposals using a fuzzy binary relations method. The keywords are searched in MEDLINE's indexed profiles of terms derived from the bibliography attributable to authors. If there is more than one author with the same name, there may be a problem in finding a suitable reviewer candidate. To deal with name ambiguity, Peer2ref automatically evaluates profiles by using authors' co-authorship information. Mittal et al. (2019) presented a strategy to address the RAP ambiguity based on fuzzy sets and expansion concepts. They first proposed an automatic keyword extraction for determining essential words in the proposals and reviewers' publications. Then they created fuzzy word-to-word graphs. Fuzzy graph centrality measures (i.e., fuzzy degree centrality, fuzzy betweenness centrality, fuzzy closeness centrality) are applied to form a fuzzy set for representing the terms and their weights for each proposal and reviewer. With fuzzy extension principle is employed to find the similarity between each proposal-reviewer pair. To enrich text representation with contextual knowledge, there are some ontologies that are used as external sources for embedding background knowledge to text documents of reviewers and proposals. For instance, WordNet (Miller, 1995) exposes many synonym associations between words. In the paper of Mittal et al. (2019), WordNet was used to find the difference between fuzzy sets of reviewers and proposals. Xue et al. (2012) presented the interval fuzzy ontology method that adds semantic matching to the RAP by using academic discipline ontology and interval-valued fuzzy sets theory in modeling the research topic of proposals and reviewers. They computed the relation between the proposal and the reviewer based on WordNet. But, since these ontologies are made by hand, their coverage is limited, and maintenance is extremely laborious. For these reasons, as a more feasible solution, Davoodi et al. (2013) built a semantic kernel with the background knowledge derived from the Wikipedia repository, which provides a knowledge graph by connecting correlating concepts. The semantic kernel is used to enrich the reviewers' profiles.

### 6.2.3 METHODS COMBINING WORD AND SEMANTIC INFORMATION

Some scholars have advocated combining word and semantic information to find the most qualified reviewers for a targeted proposal. To obtain a more accurate similarity score between reviewers and proposals, Tang et al. (2010), Tang et al. (2012) and Tan et al. (2021) used both the LDA model and SLM. Similarly, in the well-known reviewer assignment system TPMS, Charlin, and Zemel (2013) modeled the proposals and reviewers' published publications using the SLM and LDA. According to Peng et al. (2017), the scientific interests of reviewers are frequently multifaceted and may shift over time. They stated that more recent publications would better reflect the reviewers' current research interests. Therefore, the authors introduced a time-aware and topic-based assignment model. The proposed model used LDA and TF/IDF to represent reviewers and proposals for a more accurate similarity score. Combining LDA and TF/IDF was also applied by Pradhan et al. (2020) to model reviewers' research interests. Silva et al. (2014) developed a novel similarity score measurement by integrating both "exact matching" and "semantic matching" scores. The exact matching score, which is the predicted probability that the expertise of a reviewer appears on a proposal, is computed using the SLM. Semantic matching is produced from the bipartite graph's weighted edges, a reduced form of the ontology-based proposal and reviewer semantic network.

The distribution of methods focusing on semantic information of reviewers and proposals is presented in Table 2. As shown in Table 2, methodologies for topic modeling such as LSI, PLSA, and LDA are extensively used semantic-based approaches focusing on the second stage of the RAP. Because of its advantages over other topic models, the LDA model is widely utilized to identify the topics that represent reviewers and proposals.

Neural network-based language models produce outstanding results in a variety of NLP downstream tasks, including document categorization, named entity recognition, and machine translation (Romanov and Khusainova, 2019; Kalyan et al., 2021; Kapočiūtė-Dzikienė et al., 2021). Although the using such new-generation language models in calculating similarity scores is not very common, they have been applied in a few studies as of 2018. The application of neural network-based models to the calculation of similarity scores has shown that they are more effective than other semantic-based methods (Ogunleye et al., 2018; Duan et al., 2019; Yong et al., 2019). Neural network-based language models are actually in the semantic-based approach category. However, these methods are presented in a separate section, taking into account the remarkable benefits that they have brought to the calculation of similarity scores and may possibly bring in the future.

## 6.3  Neural Network-Based Language Models

The word-based models and topic models presented for the RAP are mainly frequency-based techniques. Frequency-based techniques that treat documents as a set of terms, the frequency of each term is essential, but the order is ignored. The most significant disadvantage of these methods is that they cannot encode semantic and syntactic affinities between words; therefore, the representations created are semantically weak. In addition, frequency-based techniques for large data sets frequently generate vectors with high sparsity, where the n-dimension of each document's vector representation is equal to the n-size of the document vocabulary. This problem with dimensions is often known as the curse of sparsity. Data sparsity is notably prevalent in morphologically rich languages, such as Turkish, Chinese, and Czech, due to word affixes.

| Methods | 1992 | 2007 | 2008 | 2010 | 2012 | 2013 | 2014 | 2015 | 2016 | 2017 | 2018 | 2019 | 2020 | 2021 |
|---|---|---|---|---|---|---|---|---|---|---|---|---|---|---|
| LSI | ✓ | | | | | | | | ✓ | | | ✓ | | |
| PLSA | | | ✓ | | | | | | | | | ✓ | | |
| LDA | | | | ✓ | ✓ | ✓ | ✓ | | ✓ | ✓✓ | ✓ | ✓ | ✓✓ | ✓✓✓ |
| ATM | | | ✓ | ✓ | | | | ✓✓ | | ✓ | | | | |
| APT | | ✓ | | | | ✓ | | | | | | | | |
| AST | | | | | | | | | | | ✓ | | | |
| Word2Vec | | | | | | | | | | | ✓✓ | ✓✓ | | ✓ |
| BERT | | | | | | | | | | | | ✓ | | |
| CNN | | | | | | | | | | | | ✓ | | |
| bilSTM | | | | | | | | | | | | ✓ | | |
| SciBERT | | | | | | | | | | | | ✓ | | |
| Doc2Vec | | | | | | | | | | | | | ✓ | |
| Wordnet | | | | | ✓ | | | | | | | ✓ | | |
| Semantic Kernel | | | | | | ✓ | | | | | | | | |
| Common Topic Model | | | | | | | | | | | | ✓ | | |

Table 2: Distribution of studies focusing on semantic-based methods by year.

Language models based on deep learning and neural networks have been developed to solve such issues in frequency-based techniques and automatically learn the syntactic and semantic features of the documents. Neural network-based language models use word embedding for learning word sequence distributions and the semantic representation of documents. Word embedding can be used to compute continuous-valued distributed representations of documents with low-dimensional space. Neural network-based word embedding methods are examined under two categories. The first category is called static word embedding methods or prediction-based word embedding methods, which can encode morphological and semantic features of words and consider sub-word elements (character n-grams). Word2Vec (Mikolov et al., 2013), Glove (Pennington et al., 2014), and FastText (Bojanowski et al.,2017) are the most common static-based word embedding methods. The second category is context-based word embedding methods, including models with deep architecture such as ULMFiT (Howard & Ruder, 2018), BERT (Devlin et al., 2019), GPT-2 (Radfort et al., 2019), DistillBert (Sanh et al., 2019). While static word embedding approaches represent the word as a dense vector, context-based representation methods, as the name suggests, add contextual information to the representation. With very deep architectures, they can learn the meaning of a word from its context and encode it into a vector.

In recent years, neural network-based word embedding methods have been used for many NLP tasks, including word prediction, translation, automatic speech recognition, and text similarity calculation. Regarding the task of reviewer assignment, the proposal-reviewer similarity can thus be measured by the similarity of the dense vector representation of their documents. Ogunleye et al. (2018) calculated a similarity score between the reviewer and the proposal by Word2Vec. Also, they used frequency-based models such as TF-IDF, LSI, and LDA for similarity score measurement. A comparison of the findings showed that the most favorable proposal-reviewer matches were obtained with Word2Vec. Zhao et al. (2018) developed a new classification method called Word Mover's Distance–Constructive Covering Algorithm (WMD–CCA) to solve the RAP as a classification problem. This method is based on the idea that using research field label information about reviewers and proposals can improve the quality of the assignment. First, Word2Vec was used to represent proposals and reviewers. Then minimum text distance between proposals and reviewers was measured using the word mover's distance algorithm. Finally, the field labels for the reviewer candidates and proposals were predicted according to the field relationship. This approach can learn field features through statistical learning. However, it does not allow the machine to learn field information from the right perspective and borrow from human knowledge. To help the machine to learn field relationships from the proper perspective, Duan et al. (2019) proposed a sentence pair modeling approach. This approach utilized the field association between the title and abstract of the proposals and reviewers as the supervisory information. They used six different neural network-based language models to model sentence pairs and learned the field features of the proposals and papers of reviewers through the training set, which are: BERT-CNN, Word2Vec-CNN, BERT-bilSTM, Word2Vec-CNN, random word embedding-CNN, and random word embedding-bilSTM. Finally, they predicted the similarity score between reviewers and proposals by training model. Yong et al. (2021) also utilized Word2Vec to extract hierarchical semantic representations of keywords. Following that, a rule engine is established using the knowledge graph and rule matching to create an ordered list of reviewers based on their relevance to each proposal.

Anjum et al. (2019) claimed that prior approaches, including VSM, SLM, and probabilistic topic models, were insufficient to address the vocabulary mismatch and partial topic overlap between proposals and a reviewer's expertise. Handling cases of word mismatch and partial topic overlap between proposals and reviewers, the authors proposed the common topic model method based on the Word2Vec.

In reality, a reviewer or a proposal frequently belongs to various research domains. So, the RAP was also formulated as a multilabel classification issue, with reviewers assigned based on multiple estimated labels (Zhang et al., 2019). Also, they introduced hierarchical and transparent representation, namely Hiepar, to represent the semantic information of reviewers and proposals. The Hiepar was learned from a two-level bidirectional gated recurrent unit-based network with an attention mechanism. The model is able to capture the hierarchical structure of reviewers and proposals, as well as their distinct semantic content.

Cohan et al. (2020) proposed the SPECTER model to learn independent representations of academic publications based on the current success of transformer language models pre-trained in many NLP tasks. Transformer language models, such as BERT, only consider the in-document context and do not utilize any information between documents. This hinders their capacity to learn the optimal document representations. Using a pre-trained SciBERT (Beltagy et al., 2019) model and relation signals extracted from the citation graph, the SPECTER embedding approach creates paper embeddings. Unlike currently available pre-trained language models, SPECTER can be used for downstream applications without task-specific fine-tuning. Hoang et al. (2021) developed a decision support system to select a group of reviewers for proposals automatically. In this system, the profile of reviewers and proposals are represented as a vector by using Doc2Vec (Le & Mikolov, 2014).

## 6.4 Network-Based Approaches

The text-based/content-based approach is limited to dictionary-bound relationships between words and descriptions of elements and does not discover implicit relationships. The implicit relationship is a social or academic network between reviewer and reviewer or proposal and reviewer, similar to the social networks that people build. Analyzing such relationships can strengthen the reviewer's recommendation. In this survey, approaches that benefit from social and academic relationship information are called network-based approaches.

Social network analysis is a special kind of network-based method allowing users to leverage the social structure of their informal interactions as an additional source of information. Network information sources in social network analysis can be generally cited as citation networks, collaboration networks, co-authorship networks, and numerous interconnected networks between the reviewer and reviewer, as well as the proposal and reviewer.

A citation network is a directed graph that depicts papers and their citation relationships, i.e., the connections between cited and citing articles. This graph enables the identification of direct and indirect links between documents. Some citation-related characteristics, like the number of citations within a document and the time between the publication date of the cited article and the publication date of the citing article, can also be used to differentiate the weights of links between documents and estimate how similar they are. Jin et al. (2017) and Jin et al. (2020) used a citation network to aid in the identification of potential reviewers. The proposed method is based on the following assumption: if a reviewer candidate is cited by many qualified experts in a certain field, that candidate has a high level of authority in that field. Citation actions were organized in a topical authority graph to analyze each reviewer's relative authority for each proposal. They employed topical PageRank to determine high-authority reviewers for given topics. In another research, Tran et al. (2017) sought to suggest candidates for the program committee renewal of a specific conference. According to the authors, the conference's keywords are limited, generic, and sometimes non-English. Using these keywords to search for potential reviewers in a bibliographic database yields noisy results. Therefore, this study offered a technique that does not rely solely on text. In the proposed method, a three-mode network was constructed to model the knowledge of expert candidates in a specific conference, and the candidate's proximity to the conference was determined based on the weights of their various relationships, such as conference committee membership, co-authorship, citation networks, etc.

Xu and Du (2013) argued that studies addressing the RAP and focusing on text-based information retrieval techniques fail to consolidate relevant information. To solve this problem, they proposed a three-layer network model that combines the semantic, social, and citation networks. When creating the network model, reviewers and publications were positioned as nodes, and the relationship between them was represented by connections. Semantic analysis has captured the semantic relationship of research areas. And citation analysis has been used to analyze the citation relationship between articles. The particle swarm algorithm utilized the three-information acquired by the three-layer model to provide reviewer recommendations. Rodriguez and Bollen (2008) introduced a co-authorship network-based collaborative filtering system in which the network is defined by a triple graph with vertices, edges, and weights. Each vertex represents an author in the network, each edge represents a collaborative publication by two authors, and each weight represents the strength of the relationship between any two cooperating authors. They employed a relatively advanced particle swarm technique to compute the similarity score. Silva et al. (2014) created a connectivity matrix with social networks to identify like-minded reviewers based on their knowledge-sharing behavior.

## 7. Stage Three: Optimization-Based Reviewer Assignment Problem

The third stage of the RAP, the Optimization-based Reviewer Assignment Problem (ORAP), transforms the IRRAP into an optimization problem. During this phase, the RAP is formulated based on diverse optimization objectives and constraints, and it is then solved using various optimization techniques. This section discusses mathematical models and algorithms for optimizing reviewer assignments.

In general, the ORAP approaches have a two-stage framework. In the first stage, a weighted bipartite graph is constructed between the proposal and review sets. Typically, the weight of the edge between a proposal and its reviewer shows the relevance between them. In the second stage, a final proposal-reviewer assignment is determined based on the bipartite graph to optimize an appropriate objective function under various constraints (Pradhan et al., 2020). The RAP can be seen as an advanced form of the Generalized Assignment Problem (GAP), in which two large sets of objects must be matched such that each object in one set is assigned to a limited number of objects in the other set. Since 1999, several mathematical models and algorithms have been proposed for matching a set of reviewers to a collection of proposals to maximize an overall optimization objective. Table 3 presents the distribution of current works that proposed mathematical models addressing the RAP by year. The proposed mathematical models differ in terms of optimization objectives, constraints, and strategy. In the ORAP approaches, the optimization objective is typically described as maximizing the cumulative sum of similarity scores, maximizing topic coverage, and optimizing fairness. However, it is not sufficient to simply establish a proposal-reviewer matching that optimizes objectives. The assignment must take into consideration numerous practical constraints. The constraints considered in the RAP mathematical models and their definitions are presented in Table 4. According to Table 4, although different constraints are taken into account in various studies, the following two constraints are frequently attempted to be met:

(1) each reviewer should be assigned a pre-determined number of proposals (reviewer workload)

(2) each proposal should be reviewed by a certain number of reviewers (proposal demand).

Given set $P = \{1, 2, ..., n\}$ of proposal and set $R = \{1, 2, ..., m\}$ of available reviewers. With these two commonly used constraints and the maximizing the sum of the similarity score objective, in its most general form, the RAP can be formulated as follows:

$$\max \sum_{i=1}^{n} \sum_{j=1}^{m} s_{ij} x_{ij} \qquad (7\text{-}1)$$

$$s.t. \sum_{j=1}^{m} x_{ij} = d, \quad \forall i = 1, ..., n \qquad (7\text{-}2)$$

$$\sum_{i=1}^{n} x_{ij} = w, \quad \forall j = 1, ..., m \qquad (7\text{-}3)$$

where $x_{ij} = \begin{cases} 1, & \text{if reviewer } j \text{ is assigned to proposal } i \\ 0, & \text{otherwise} \end{cases}$

In equation 7-1, $s_{ij}$ denotes the similarity score between a proposal $i$ and a reviewer $j$. Equations 7-2 and 7-3 express the constraints on the number of reviews required per proposal ($d$) and the workload of reviewers ($w$), respectively.

This survey discusses the ORAP approaches under the following three subsections. The first subsection discusses the different optimization objectives of the problem. The second one presents optimization strategies for special objectives. The last one mentions other issues related to problem formulation. Furthermore, a detailed summary of the studies addressing the RAP from an optimization perspective is given in Appendix C. In Appendix C, you can find the proposed mathematical models and algorithms for solving these models. You can also find the different performance comparison strategies and metrics used to measure how well the proposed approaches work.

## 7.1 Optimization Objectives of the Problem

The studies addressing the RAP from an optimization perspective proposed different mathematical models and algorithms to achieve determined optimization objectives. Based on the optimization objectives, this survey divides studies into three categories: maximizing the cumulative sum of similarity scores, maximizing topic coverage, and optimizing fairness. Each category is discussed separately in the following sections.

### 7.1.1 MAXIMIZING CUMULATIVE SUM OF SIMILARITY SCORES

The most common optimization objective used in studies is to maximize the cumulative sum of similarity scores for all matched proposal-reviewer pairs (Cook et al., 2005; Goldsmith & Sloan, 2007; Taylor, 2008; Kolasa & Krol, 2011; Charlin et al., 2012; Charlin & Zemel, 2013; Long et al., 2013; Li & Hou, 2016). These studies assumed the availability of a list of submitted proposals, reviewers, and a proposal-reviewer similarity score matrix. They believed the most qualified reviewer would be allocated to each proposal when the total similarity score reached its maximum value. Several well-known conference management tools, such as TPMS (Charlin & Zemel, 2013), EasyChair (easychair.org), and HotCRP (hotcrp.com), have implemented the maximization of the total similarity optimization strategy.

To maximize the sum of similarity scores, most researchers have structured the RAP as an integer linear programming (ILP) formulation (Guervós & Valdivieso, 2004; Janak et al., 2006; Li et al., 2008; Taylor et al., 2008; Conry et al., 2009; Xu et al., 2010; Chen et al., 2012; Charlin et al., 2012; Li et al., 2013; Charlin &

Zemel, 2013; Silva et al., 2014; Li & Hou, 2016; Ogunleyu et al., 2017; Jin et al., 2018b; Yang et al., 2020; Kat, 2021).

Some researchers proposed a formulation known as mixed integer programming (MIP), which includes an additional set of continuous variables into the objective function for constraint relaxation. (Janak et al., 2006; Garg et al., 2010; Wang et al., 2013; Lian et al., 2018; Yeşilçimen &Yıldırım, 2019; Leyton-Brown et al., 2022).

When the number of reviewers and proposals is small, exact algorithms or typical linear programming solvers (e.g., CPLEX) can produce a proposal-reviewer assignment solution that optimizes the cumulative similarity score (Janak et al., 2006; Taylor et al., 2008; Charlin et al., 2012; Ogunleyu et al., 2017; Jin et al., 2018b). Due to the dual difficulties of a large volume of proposals and reviewers and limited time, the problem cannot be addressed efficiently by exact algorithms in an acceptable amount of time and memory. Therefore, numerous researchers have investigated the use of approximation or heuristic algorithms for the problem of allocating reviewers, such as the greedy algorithm (Cook et al., 2005; Karimzadehgan & Zhai, 2009, 2012; Long et al., 2013; Guo et al., 2018; Boehmer et al., 2021), two-phase stochastic-biased greedy algorithm (Wang et al., 2013), stage deepening greedy algorithm (Kou et al., 2015a, 2015b; Mirzaei et al., 2019), weighted-matrix factorization based greedy algorithm (Pradhan et al., 2020), greedy reviewer round-robin algorithm (Payan & Zick, 2022), ant colony algorithm (Li et al., 2008), random split (Jecmen et al., 2021), row/column generation (Leyton-Brown et al., 2022), genetic algorithm (Li et al., 2007; Xu et al., 2010; Chen et al., 2012; Li et al., 2013), particle swarm optimization algorithm (Yang et al., 2020), hill-climbing algorithm (Hettich & Pazzani, 2006), OBH algorithm (Yeşilçimen & Yıldırım, 2019), PeerReview4All (Stelmakh et al., 2019) and FairFlow (Kobren et al., 2019). Some researchers have proposed hybrid algorithms, which are a combination of more than one algorithm and are used to eliminate the drawbacks of single algorithms (Yang et al., 2020). For instance, Guervós and Valdivieso (2004) attempted to solve the RAP by combining greedy and evolutionary algorithms. The proposed hybrid algorithm produced a more precise proposal-reviewer match in less time than the greedy algorithm alone. Xu et al. (2010) introduced a hybrid algorithm based on greedy randomized adaptive search processes and genetic algorithms to overcome the limitations of the genetic algorithm and make it more effective by combining it with the local search strategy. Kolasa and Krol (2011) combined genetic algorithm and ant colony optimization (GA&ACO) to achieve feasible assignments and prevent premature convergence. The tabu search and genetic algorithm (TS&GA) hybrid algorithm were also introduced in the same study for similar purposes. Experiments have demonstrated that hybrid algorithms perform poorly in terms of time, despite giving more precise fitness-based assignments than genetic algorithms or ant colony optimization alone.

The Network Flow Model works well for the RAP when proposals and reviewers are represented as nodes in a bipartite graph. Each edge in the graph corresponds to cost and capacity. The cost is determined by the relevance between each reviewer and each proposal. By adding a source and a sink node and defining what each arc can do, it is possible to meet capacity limits like the number of proposals and the number of reviewers. Researchers have presented different approaches based on network flow theory, such as the maximum cost of flow problem (Goldsmith & Sloan, 2007), minimum cost of flow problem (Huang et al., 2010), minimum convex cost flow problem (Tang et al., 2010; Tang et al., 2012) and equivalent minimum convex cost flow problem (Yan et al., 2017).

| Mathematical Model | 1999 | 2004 | 2005 | 2006 | 2007 | 2008 | 2009 | 2010 | 2011 | 2012 | 2013 | 2014 | 2015 | 2016 | 2017 | 2018 | 2019 | 2020 | 2021 | 2022 |
|---|---|---|---|---|---|---|---|---|---|---|---|---|---|---|---|---|---|---|---|---|
| Generalized Assignment Problem (ILP) | | ✓ | | ✓ | ✓✓ | ✓ | ✓ | ✓ | | ✓✓✓✓ | ✓✓ | ✓✓ | ✓ | | ✓✓✓ | | ✓✓ | ✓✓ | ✓✓ | |
| Mixed ILP | | | | ✓ | | | | | | | | ✓ | | | | ✓ | ✓ | | | ✓ |
| Minimum Convex Cost Flow Problem | | | | | | | ✓ | ✓ | | | | | | | | | | | | |
| Pairwise Constrained Problem | | | | | | | | | | | | ✓ | | | | | | ✓ | | |
| Network Flow Theory | | | | ✓ | | | | | | | | | ✓ | | | | | | | |
| Constrained Multi-Aspect Committee Review Assignment (CMACRA) | | | | | | | ✓ | | | | | | | | | | ✓ | | | |
| Weighted Coverage Group-Based Reviewer Assignment Problem | | | | | | | | | | | | | ✓✓ | | | | | | | |
| Capacitated Bottleneck Transshipment Problem | ✓ | | | | | | | | | | | | | | | | | | | |
| Set Covering Problem | | | ✓ | | | | | | | | | | | | | | | | | |
| Minimum Cost Flow Problem | | | | | | | | ✓ | | | | | | | | | | | | |
| Maximum Topic Coverage Paper Reviewer Assignment (MaxTC-PRA) | | | | | | | | | | | ✓ | | | | | | | | | |
| Fuzzy LP | | | | | | | | | | | | ✓ | | | | | | | | |
| Capacitated Transportation Problem | | | | | | | | | | | | | | ✓ | | | | | | |
| Equivalent Minimum Convex Cost Flow Problem | | | | | | | | | | | | | | | ✓ | | | | | |
| Equilibrium Multi-Job Assignment | | | | | | | | | | | | | | | | | | ✓ | | |
| Strategy Proof | | | | | | | | | | | | | | | | | | ✓ | | ✓ |

| | | | | | | | | | | | | | | | |
|---|---|---|---|---|---|---|---|---|---|---|---|---|---|---|---|
| via Partitioning Assignment | | | | | | | | | | | | | | | |
| Weighted z-Cycle-Free Peer Reviewing | | | | | | | | | | | ✓ | | ✓ | | |
| Two Stage Paper Assignment Problem | | | | | | | | | | | | | | ✓ | |

Table 3: Distribution of mathematical models by year.

Cook et al. (2005) offered an alternative strategy for the RAP. They modeled this problem as the Set Covering Problem, whose main goal is maximizing the sum of the number of joint panelists evaluating each pair. Lian et al. (2018) conceived the RAP as two-sided matchmaking with preferences of one side (reviewers) over the other side (proposals). Considering that both parties have capacity constraints, the problem is modeled as MIP. Unlike other studies, upper and lower capacity limits were used for both sides. This study uses order weighted averages (OWAs) in the proposed flow network algorithm for the RAP, inspired by studies in multi-criteria decision-making and social choice theory. Pradhan et al. (2020) formulated the RAP as an equilibrium model of a multi-job assignment problem. In this case, the job corresponds to the proposal and the worker to a reviewer. They aimed to maximize the total profit by maximizing the topic similarity and minimizing the COIs under the balanced workload distribution of the reviewers. To solve the model, they presented a greedy algorithm based on weighted matrix factorization, in which total profits are incrementally increased by searching for an assignment where the difference between the maximum and minimum total profit is highest for each reviewer.

Some commonly used decision science techniques, such as fuzzy optimization and analytic hierarchy process (AHP), could also be employed to facilitate the RAP. For example, Daş and Gökçen (2014) handled the RAP in terms of inherent uncertainty. They introduced a fuzzy ILP model aiming to maximize the similarity score between proposals and reviewer groups while minimizing the cost of generating a panel of reviewers under crisp constraints. The cost of the panel is composed of fees paid to reviewers for evaluation and expenses such as travel and food incurred by reviewers. They added a constraint to the model to ensure that the cost did not exceed a certain budget. Tayal et al. (2014) introduced another fuzzy optimization approach for solving the RAP, which optimizes the equality score between the type-2 fuzzy expertise sets of reviewers and the fuzzy keyword sets of proposals, subjected to a set of relevant constraints. This research presented an integrated solution that encompasses all aspects of expertise modeling. Xue et al. (2012) modified the mathematical model of Tang et al. (2012) using the interval fuzzy ontology method to compute similarity scores. Li and Watanabe (2013) established an integrated model for reviewer suggestions that consider both reviewers' relevance and expertise. The expertise level and relevance score are calculated using an AHP-based method.

### 7.1.2 Maximizing Topic Coverage

The cumulative expertise of all the assigned reviewers to a proposal should cover all aspects of the proposal. Considering a proposal as a single topic without regard to its multiple aspects will result in inadequate reviewer assignments, especially for proposals with more than one subtopic. Therefore, some researchers considered maximizing topic coverage as a global objective and proposed different approaches to maintain comprehensive coverage (Karimzadehgan & Zhai, 2009, 2012; Tang et al., 2010; Xue et al., 2012; Tang et al., 2012; Long et al., 2013; Tayal et al., 2014; Kale et al., 2015). They represented the reviewers' expertise and the proposals' content into two sets of topics or assumed that these two sets were available. These topic sets may be derived from the list of subject keywords generally provided by a conference, journal, or grant organization, or they may be automatically discovered using topic models. Then they evaluated the quality of the reviewer assignment scheme using the different coverage ratios of topics.

Long et al. (2013) presented the Maximum Topic Coverage Paper Reviewer Assignment (MaxTC-PRA) problem in which the quality of the assignment of a group of reviewers to a proposal is measured by the set-coverage ratio. It is the ratio of the number of proposal topics that the reviewer group covered to the total number of proposal topics. Tang et al. (2010) and Tang et al. (2012) proposed a constrained-based optimization problem considering ''multi-topic coverage'' proposal-reviewer matching. They incorporated two topic coverage measures into the proposed model as constraints. The first one is coverage which satisfies that the reviewer group assigned to each proposal should be able to cover as many aspects of the proposal as possible. The formulation of coverage is presented in equation 7-4. The second one is confidence which means that the competence of each reviewer assigned to a proposal should encompass as many aspects of the proposal as possible. Confidence can be computed by equation 7-5. To solve the proposed optimization framework, they transformed the problem into a convex-cost network flow problem and provided an efficient algorithm that ensures an optimal solution.

$$Coverage(q_i) = \frac{\left| T(q_i) \cap \cup_{v_j \in V(q_i)} T(v_j) \right|}{|T(q_i)|} \tag{7-4}$$

$$Confidence(q_i) = \frac{1}{m} \sum_{v_j \in V(q_i)} \frac{\left| T(q_i) \cap T(v_j) \right|}{\left| T(q_i) \cap \cup_{v_j \in V(q_i)} T(v_j) \right|} \tag{7-5}$$

where, the relevant topics of the proposal $(q_i)$ and reviewer $(v_j)$ are represented as $T(q_i)$ and $T(v_j)$ demonstrating the most related aspects to the proposal and reviewer, respectively. $V(q_i)$ is set of the reviewer candidates.

These proposed strategies improve the aspect coverage of the review assignments. However, in these approaches, reviewers are assigned to each proposal independently, without considering the entire committee. This makes it difficult to balance the review workload among a group of reviewers and may lead to assigning an excessive number of proposals to a reviewer with competence in a popular field.

Also, while some proposals are evaluated by knowledgeable reviewers, it may be inevitable that others will be assigned to reviewers with limited expertise. Unlike the approaches that assign reviewers to proposals independently, some scholars defined a multi-aspect group-based assignment.

Karimzadehgan and Zhai (2009, 2012) introduced the Constrained Multi-Aspect Committee Review Assignment (CMACRA) problem, in which the aim is to assign a group of reviewers to a set of proposals based on multi-aspect proposal-reviewer matching under reviewer workload and proposal demand constraints. They cast the CMACRA problem as an ILP. They used coverage and confidence measures to determine assignment quality. Finding the optimal values for coverage and confidence is typically NP-hard (Zhai et al., 2003); hence they employed greedy techniques to obtain an approximation.

Based on the weighted coverage degree, they defined Weighted-coverage Group-based Reviewer Assignment Problem (WGRAP) model to maximize the total weighted coverage ratings per topic. Additionally, Kou et al. (2015a, 2015b) presented an approximation algorithm called stage deepening greedy algorithm that achieved a 1/2-approximation ratio. Another group-based multi-aspect proposal-reviewer assignment formulation was proposed by Mirzaei et al. (2019), which was called "Multi-Aspect Review-Team Assignment using Latent Research Areas (MARTA-LRA). This approach used the latent research areas, which were expressed by concatenating the terms of the reviewers' publications, to determine the importance of each topic of proposals and reviewers.

Kale et al. (2015) have added a new dimension to the maximum topic coverage optimization strategy by incorporating the reviewers' freshness. The impact factor of the journals in which the reviewer's papers were published was utilized to determine the freshness of the reviewer. The Maximum Topic Coverage Paper-Reviewer Assignment Algorithm based on the greedy strategy was introduced to ensure that reviewers looked at as many different proposal topics as possible.

### 7.1.3 OPTIMIZING FAIRNESS

Even though maximizing the overall sum of similarity scores or optimizing topic coverage can produce an accurate reviewer assignment generally, it can lead to unfairness for some proposals. The main reason is that the global objective can be optimized if more qualified reviewers are given to one proposal at the expense of the other. In this approach, some proposals may be assigned to reviewers who lack expertise in the subject areas of these proposals, while others may be assigned to competent reviewers. Kobren et al. (2019) analyzed data from CVPR'17,18 conferences with several thousand papers. The analysis showed that at least one proposal in the assignment results with this approach had a similarity score of zero for all reviewers. For optimizing fairness, no reviewer should benefit at the expense of others.

Hartvigsen et al. (1999) partially addressed the issue of fairness by maximizing total expertise under a constraint that ensures each proposal has at least one reviewer with expertise above a certain threshold. This method assigned one highly specialized reviewer to each proposal but failed to prevent discrimination against some proposals in the assignment of the remaining reviewers. Recent studies have attempted to overcome these problems by (a) imposing strict constraints on the minimum similarity of valid proposal-reviewer matches or (b) optimizing the sum of the similarity of a proposal in the worst case.

To achieve fairness, Garg et al. (2010) aimed to maximize the utility of the worst-off proposal. Taking into account the lexicographic preferences of the reviewers, they introduced the lexi-min and rank-maximal assignment notions based on the max-min fairness objective, which is shown in equation 7-6. They believed that the assignment is fair if the sum of similarity across all proposals is minimum under the coverage and workload constraints. They proved that finding a lexi-min or an optimal egalitarian assignment if there are more than two equivalence classes is NP-hard.

$$(g, p) = max_M \, min_j \, \sigma_j(M) \tag{7-6}$$

where $\sigma_j(M)$ is the signature of reviewer $j$ under assignment $M$. Signatures are compared by either the weighted preference or lexicographic rank.

The objective of fairness addressed in the study of Stelmakh et al. (2019) is to maximize the review quality of the most disadvantaged proposal rather than to maximize the total review quality of all proposals. Inspired by the notion of max-min fairness, they aim to find feasible assignment $A$ to maximize the fairness objective $\Gamma^S$ for given similarity matrix $S$. $R_A(i)$ denotes the set of reviewers who review proposal $i$ under an assignment $A$. The formulation of the fairness objective is presented in equation 7-7.

$$\Gamma^S(A) = min_{i \in [n]} \sum_{j \in R_A(i)} s_{ij} \tag{7-7}$$

Unlike various approaches that optimize assignments for specific deterministic targets for the RAP, this study also examines assignments for statistical accuracy. They introduced the PeerReview4All algorithm based on the incremental max-flow technique to meet the essential concerns of fairness and statistical accuracy. The algorithm achieved an optimal solution within a constant factor in terms of the maximum-minimum fairness objective. In addition, the PeerReview4All algorithm provides robust statistical assurances for correctly identifying the best proposals that should be accepted.

Proposed algorithms that maximize the sum of similarities for the worst-off proposals do produce more fair assignments, but they have some drawbacks: (i) constraining the minimum allowable similarity can often make the problem infeasible, as there may not be any match that provides adequate coverage to all proposals subject to the threshold. (ii) They do not impose load balance among reviewers and may therefore yield matchings in which reviewers are assigned to wildly diverse quantities of proposals. Kobren et al. (2019) presented a new model that considers local fairness to eliminate these issues. This model is formulated as an integer linear program that optimizes the global objective, includes both upper and lower limit constraints to balance the workload of the reviewers, and contains local fairness constraints to ensure that each proposal is collectively assigned to a group of reviewers with sufficient expertise. The local fairness constraint can be formulated, as shown in equation 7-8, which states that the overall similarity score for each proposal must be at least $T$.

$$\sum_{i=1}^{n} x_{ij} s_{ij} \geq T, \quad \forall j = 1,2, \ldots, m \tag{7-8}$$

Since the proposed model is NP-hard, two algorithms are introduced in its solution. The first one is FairIr, which provides a fair assignment that maximizes the global objective and satisfies local fairness and workload constraints. The second algorithm, FairFlow, does not guarantee to produce fair matches. However, it can be faster than FairIR and PeerReview4All.

Payan and Zick (2021) formulated the RAP as a kind of fair allocation problem for the fair distribution of the reviewer's quality to the proposals. In this study, envy is the main criterion emphasized for fairness, which means if one proposal prefers reviewers assigned to another proposal more than its own, it will envy the other. Because of the impossibility of obtaining envy-free assignments for indivisible items, they focused on the criterion of being envy-free up to one item (EF1). In conventional fair allocation circumstances, the well-known round-robin method generates EF1 assignments by establishing an order for agents and allowing them to select one item at a time. Though, due to the restriction that no proposal can be examined twice by the same reviewer, round-robin is not EF1. To overcome these obstacles, they proposed the Reviewer Round Robin mechanism, which is an extension of the traditional round-robin technique.

| Constraints | Description | | References* |
|---|---|---|---|
| Reviewer workload | Each reviewer should get no more than a certain number of proposals. | | 1-3, 5-21, 23-30, 32, 34-37, 39-52 |
| Proposal demand | Each proposal should be assigned to a certain number of reviewers. | | 1-2, 4-11, 12-17, 19, 21-23, 25-30, 34-37, 39, 41-46, 48-52 |
| COIs avoidance | A reviewer should not be assigned to a proposal if there is any sort of conflict of interest between them. | affiliation relationship | 5, 19, 29-30, 42, 52 |
| | | co-author relationship | 5, 12-14, 19, 26, 29, 37, 45, 51, 52 |
| | | advisor-advisee relationship | 15, 19, 21, 29 |
| | | competitor relationship | 19, 30, 46 |
| | | project-level collaborations | 26 |
| Strategyproofness | No reviewer can change the outcome of a proposal with which they have a conflict of interest. | | 42, 44 |
| Torpedo reviewing | The reviewer assignment should guarantee that proposals are unlikely to have been torpedo-reviewed. The term "torpedo reviewing" refers to the situation in which reviewers purposefully attempt to be assigned to a proposal they dislike to reject. | | 35 |
| Cycle/Loop-free constraint | A review assignment without the following type of collusion ring should be located: A cycle of reviewers reviews a proposal provided by the next reviewer, resulting in a review cycle in which each reviewer provides good ratings. | | 35, 39, 42, 45 |
| Expertise of reviewers | The reviewers assigned to a proposal should have the necessary knowledge to evaluate it. | | 16, 26, 40 |
| Authority balance | The authority of reviewers should be balanced for each proposal. | | 13-14 |
| Reviewer de-anonymization | No reviewer's identity can be determined with high certainty for each proposal. | | 35 |
| Seniority | Experienced reviewers should be distributed relatively across proposals | | 45 |
| Cost of the panel | A panel's cost should not exceed a pre-determined budget. | | 25 |
| Size of the panel | The size of a panel should not exceed a pre-determined number. | | 25 |
| Geographic Diversity | The reviewers assigned to the same proposal must not be from the same geographical region. | | 45 |
| Position | If several positions of reviewers can be assigned to each proposal, (i) each proposal must be evaluated only once by each position, and (ii) each reviewer must serve the same number of times in each of the different positions. | | 4 |

* (1) Hartvigsen et al., 1999 (2) Guerv´os & Valdivieso, 2004 (3) Cook et al., 2005 (4) Janak et al., 2006; (5) Sun et al., 2007 (6) Li et al., 2007 (7) Li et al., 2008 (8) Karimzadehgan & Zhai, 2009 (9) Conry et al., 2009 (10) Xu et al., 2010 (11) Garg et al., 2010 (12) Liu and Hong, 2010 (13) Tang et al., 2010 (14) Tang et al., 2012 (15) Chen et al., 2012  (16) Karimzadehgan & Zhai, 2012 (17) Xue et al., 2012 (18) Charlin et al., 2012 (19) Long et al., 2013 (20) Tayal et al., 2014 (21) Wang et al., 2013 (22) Li et al., 2013 (23) Charlin & Zemel, 2013 (24) Cechlarova et al., 2014 (25) Daş & Gökçen, 2014 (26) Silva et al., 2014 (27) Kale et al., 2015 (28) Kou et al., 2015a, 2015b (29) Liu et al., 2016 (30) Yue et al., 2017 (31) Ogunleyu et al., 2017 (32) Yeşilçimen & Yıldırım, 2019 (33) Mirzaei et al., 2019 (34) Yang et al., 2020 (35) Jecmen et al., 2020 (36) Jin et al., 2018b (37) Pradhan et al., 2020 (38) Kat, 2021 (39) Boehmer et al., 2021 (40) Kobren et al., 2019 (41) Dhull et al., 2022 (42) Guo et al., 2018 (43) Xu et al., 2019 (44) Lian et al., 2018 (45) Leyton-Brown et al., 2022 (46) Yan et al., 2017 (47) Jecmen et al., 2021 (48) Stelmakh et al., 2019 (49) Payan & Zick, 2021, (50) Jin et al., 2017, (51) Li & Hou, 2016, (52) Hoang et al., 2021

Table 4: Constraints addressed in the RAP mathematical models and their definitions.

## 7.2 Optimization Strategies for Special Objectives

Avoiding malicious behavior has been handled with different optimization strategies in the literature. In this section, studies on avoiding malicious behavior are discussed. Peer review is an effective and widely used method for evaluating proposals' feasibility and quality. It is also a competitive procedure, meaning that the outcome of a proposal is influenced by the evaluations of other proposals. Since assigned reviewers directly create and/or contribute to the review results, reviewer assignment has been one of the most critical tasks in the review process. Due to intense competition, peer review is vulnerable to strategic manipulations of reviewers.

In the process of assigning reviewers, it is common to ask reviewers about their expertise and interests. This opens the door to manipulation since reviewers could be dishonest about their knowledge and interests. For example, reviewers usually state their preferences for proposals they want to review before running an assignment algorithm to determine which proposal they will be assigned to. In such a case, the reviewer may deliberately determine his preferences to be assigned to a particular proposal. The problem of manipulation is not limited to the bidding system, as a lot of information used to determine paper assignments can potentially be manipulated, such as the reviewers' declared area of expertise or the list of publications of reviewers. Moreover, unethical reviewers may collaborate informally with other reviewers who have submitted a proposal at the same conference or grant organization. Based on this agreement, these reviewers seek to be assigned to each other's proposals and receive a positive review in exchange for some external reward.

As stated by Thurner and Hanel (2011), even a small number of self-centered, strategic reviewers can significantly diminish the quality of the review process. Hence, finding a fair reviewer assignment to ensure that no proposal is discriminated against has been addressed by many researchers. This section discusses malicious behaviors that affect fairness in reviewer assignments and the various solutions to avoid them in the following three headings.

### 7.2.1 CYCLE/LOOP PREVENTION

Many people have two roles in conferences and grant organizations. They are applicants who submit proposals and are also designated as reviewers to evaluate the proposals. Consequently, applicants and reviewers working in related fields are more likely to have academic relationships that extend beyond COIs. In the worst-case scenario, applicants of two separate proposals are assigned to review each other's proposals as reviewers. They may work together and award higher ratings than they should. Some researchers called this behavior "collusion rings," which are groups of reviewers who unethically evaluate and support one another (Littmann, 2021). Such reviewer-proposal loops violate anonymity and threaten the fairness and legitimacy of the peer review process and should therefore be avoided in automatic reviewer assignments.

Guo et al. (2018) tried to avoid collusion rings by defining loop constraints to guarantee that the assignment does not include k-cycle reviewers. Let's consider this as a two-cycle ring; if reviewer A is assigned to review the proposal of reviewer B, reviewer B cannot be assigned to A's proposal as a reviewer. They developed an algorithm for detecting cycles. This algorithm detects cycles using topological sorting and the algorithm for Depth-First Search. They formulated the RAP as follows in order to maximize the sum of the similarity scores between proposals and reviewers (equation 7-9) under the three constraints: proposal demand (equation 7-10), reviewer workload (equation 7-11), and k-loop free constraint (equation 7-12). They claimed that as the length of the loop increased, the hardness of collaborative cheating increased. Therefore, they allowed large loops in the proposed model only if their length was larger than a predefined threshold $k$.

$$max_{x(i,j)} \sum_j \sum_i s_{ij} x_{ij} \tag{7-9}$$

$$s.t. \sum_j x_{ij} = d, \quad \forall i \in B \cup C \tag{7-10}$$

$$\sum_i x_{ij} \leq w, \quad \forall j \in A \cup B \tag{7-11}$$

$$l(g) \geq k, \ \forall g \in G \tag{7-12}$$

$$x_{ij} \in \{0,1\}, \ \forall j \in A \cup B, \ \forall i \in B \cup C \tag{7-13}$$

where $s_{ij}$ is the similarity score between reviewer $j$ and proposal $i$. As shown in equation 7-13 $x_{ij}$ is a binary assignment variable and $x_{ij} = 0$ if reviewer $j$ has COIs with proposal $i$. $w$ is the maximum reviewer workload capacity, and $d$ is the number of required reviewers for each proposal. They represented the reviewer assignment by the directed graph $G(V, E)$. In this directed graph, (i) nodes ($V$) are categorized into three classes: $A$, $B$, and $C$. Node $A$ is the reviewer-only node with an in-degree of zero. Node B represents people who were both the applicants and reviewers. Node $C$ was the applicants-only node with an out-degree of zero. (ii) Edges ($E$) denote the assignments between the reviewer and applicant nodes. (iii) $\{r, p\}$ indicates that the proposal submitted by $p$ is reviewed by $r$. (iv) $l(g)$ represents the length of the loop in subgraph $g$ where $g \in G$.

Closer to Guo et al. (2018) setting, Boehmer et al. (2021) also consider the computation of cycle-free review assignments. They proposed a weighted cycle-free reviewer assignment problem and developed greedy algorithms to solve this problem. They conducted theoretical analysis for proposed algorithms and demonstrated that the heuristic always finds a cycle-free assignment with a minimal loss in quality.

Leyton-Brown et al. (2022) introduced "no 2-cycles" constraints to prevent instances in which two reviewers bid favorably on one other's proposals and were assigned to assess these proposals. They proposed slack variables $s_{jj'ii'}^{cy} \geq 0$ shown in equation 7-14 for each bidding cycle and added a penalty variable $p^{cy}$ shown in equation 7-15 for each reviewer assignment associated with the bidding cycle.

$$x_{ij} \leq s_{jj'ii'}^{cy} \quad \forall j, j', i, i' \in CY \tag{7-14}$$

$$Objective^{cy} = \sum_{j,j',i,i' \in CY} p^{cy} s^{cy}_{jj'ii'} \qquad (7\text{-}15)$$

where reviewer $j$ bids favorably on proposal $i$ submitted by reviewer $j'$ and $j'$ bids favorably on proposal $i'$ submitted by reviewer $j$. Binary variables $x_{ij}$ denotes whether proposal $i$ is assigned to reviewer $j$. $CY$ represents a set of cycles.

These proposed cycle-free assignment approaches can avoid collusion rings for the same venue. However, this prevention can be overwhelmed by unethical people who avoid creating a cycle, such as when a reviewer assists an applicant at a conference, and the applicant returns the favor elsewhere.

### 7.2.2 TORPEDO REVIEWING AVOIDANCE

The assignment can also be manipulated through "torpedo reviewing," in which a reviewer intentionally wants to be assigned to a proposal to reject it, potentially because it is a rival proposal or because the reviewer dislikes the proposal's subject area. For instance, a reviewer may assign low ratings to a competing proposal in the hopes that his/her proposal will be accepted, thus diminishing the likelihood of the rival proposal being chosen.

In order to avoid or mitigate this type of manipulation, Dhull et al. (2022) proposed a maximum similarity strategyproof reviewer assignment, in which no reviewer can alter their evaluation to better the outcome of their own proposal. They used the partitioning method, which was introduced by Alon et al. (2011), to ensure strategyproofness. With the partitioning approach, proposals are divided into subgroups, and no reviewer is allocated a proposal from the same subgroup as their own. The reviewer assessments are then combined individually for each subgroup, so no reviewer's ratings can impact the final result of their proposal. They proposed polynomial-time algorithms and provided theoretical guarantees on their algorithm's assignment quality. Xu et al. (2020) also dealt with the assignment of reviewers to proposals that maximize the similarity score under the constraints of strategyproofness. Dhull et al. (2022) assumed that each reviewer submits a proposal and that each submission is written by a reviewer. In contrast, Xu et al. (2020) considered arbitrary authorships in which each proposal may have many authors, and each reviewer may have submitted multiple proposals. Also, they developed an algorithm based on the partitioning method, but it does not provide a theoretical guarantee of the quality of the assignment.

These three works examined only the reviewers' motivations to have their own proposals accepted. Jecmen et al. (2020) thought of other possible motivations for a reviewer to provide untruthful reviews, such as a dislike for the research field or disincentives caused by applicant-reviewer collaboration. Jecmen et al. (2020) proposed randomized reviewer-paper assignments by optimizing the sum of similarity but with some randomness to prevent torpedo reviewing, untruthful favorable reviews, and reviewer reviews de-anonymization. The suggested method aims to ensure that this proposal-reviewer pair has only a small chance of being allocated regardless of reviewer preferences and similarities.

### 7.2.3 CONFLICT OF INTEREST AVOIDANCE

Another important requirement is that "no reviewer should be assigned to a proposal with which they are in any kind of conflict". Typically, potential COIs between the proposal and the reviewer are detected at the second stage of the RAP procedure. In stage 3, strategies for avoiding these discovered COs are incorporated into the RAP mathematical models. This survey discusses COIs detection methods in the "5.1 Conflict of Interest Detection" section. This section presents how COIs avoidance strategies are integrated into the RAP models. To ensure that there is no COIs in the assignment, it is common to set a binary decision variable $x_{ij}$ to zero in the objective function (Guerv´os & Valdivieso, 2004; Janak et al., 2006; Karimzadehgan & Zhai, 2009, 2012; Tang et al., 2010; Tang et al., 2012; Chen et al., 2012; Xue et al., 2012). Sun et al. (2008) prevented injustice by setting the level of expertise to zero for reviewers with COIs. Long et al. (2013) created a COIs set $\mathcal{C}$ that includes all potential conflicts between proposals and reviewers. The proposed algorithm of that study searched for set $A$ that contains pair of proposal-reviewer assignments and does not intersect with set $\mathcal{C}$. To remove COIs, Silva et al. (2014) identified an indicator variable $c_{ij}$ and added to the objective function. They set $c_{ij} = 0$ to show that the assignment is impossible. Otherwise, $c_{ij}$ is set to 1. Wang et al. (2013) also used the indicator variable $c_{ij}$ and they generated a constraint function based on this variable to ensure that there were no COIs.

Yan et al. (2017) proposed a new formulation called Minimum COIs Paper Reviewer Assignment (MinCOI-PRA), which aims to maximize topic similarity and minimize COIs. They defined $COI(A)$ of each assignment as in equation 7-16. In this formulation, the COIs among researchers (author to reviewer) are represented as $COI_V$ and COIs of researchers' institutions are represented as $COI_D$.

$$COI(A) = \gamma \sum_{i\epsilon P, j\epsilon R} COI_V\left(F(i), j\right) + (1 - \gamma) \sum_{i\epsilon P, j\epsilon R} COI_D\left(D\big(F(i)\big), D(j)\right) \qquad (7\text{-}16)$$

where $\gamma \epsilon [0,1]$ is a coupling parameter used to balance $COI_V$ and $COI_D$. Each proposal's applicant is denoted as $F(i)$ while their affiliate institutions are denoted as $D\big(F(i)\big)$. Each reviewer's related institution is denoted as $D(j)$.

Recent research by Pradhan et al. (2020) formulized the RAP as a maximization type equilibrium multi-job assignment problem in which the profit is incrementally optimized by maximizing topic relevance and minimizing COIs for each reviewer. To meet minimum COIs requirements between reviewers and proposals, they developed COIs value that is inversely proportional to co-authorship distance.

## 7.3 Other Issues Related to Problem Formulation

Apart from establishing and optimizing the mathematical model in ORAP approaches, some studies deal with the structure of the problem. In this survey, these studies are discussed under two headings: focusing on grouping strategies instead of assigning proposals one-on-one with reviewers and a two-stage evaluation strategy that gradually performs the assignment process when many proposals are received.

### 7.3.1 GROUPING STRATEGY

Most of the existing approaches rely on the same strategy aiming to identify suitable reviewers for each proposal (as shown in Figure 11A). This strategy may lead to unfair review outcomes due to the different subject-area experiences of reviewers assigned to similar topic proposals (Wang et al., 2013). For instance, a proposal regarding sustainable supply chain management could be reviewed by two experts in environmental science and four experts in supply chain management. However, another proposal on the same subject may be evaluated by three experts in ecological science and two experts in supply chain management. Also, when many proposals are submitted and numerous qualified reviewers are available, choosing suitable reviewers for each proposal takes a lot of time (Xu et al., 2010).

For the evaluation of proposals on similar subjects by similar reviewers, Xu et al. (2010) proposed a new reviewer assignment strategy (as shown in Figure 11B) based on the idea of "grouping". According to this strategy, proposals are first grouped by research areas, and then a set of reviewers is assigned to each proposal in the group. Inspired form the grouping idea, Daş and Gökçen (2014) proposed a novel assignment strategy in which, instead of allocating a proposal group to a reviewer, they determine the best reviewers from the pool of potential reviewers for each proposal in the group. To maximize the similarity score between proposal and reviewer groups while minimizing the cost of producing reviewer groups under strict constraints, they developed a fuzzy ILP model. Wang et al. (2013) defined the group-to-group assignment strategy (as shown in Figure 11C), in which proposals and reviewers are separated into distinct groups according to similarities, and then each group of proposals is assigned to a distinct group of reviewers. The researchers formulated the RAP as a multipurpose mixed integer programming problem and proposed a two-phase stochastic-biased greedy algorithm to solve it.

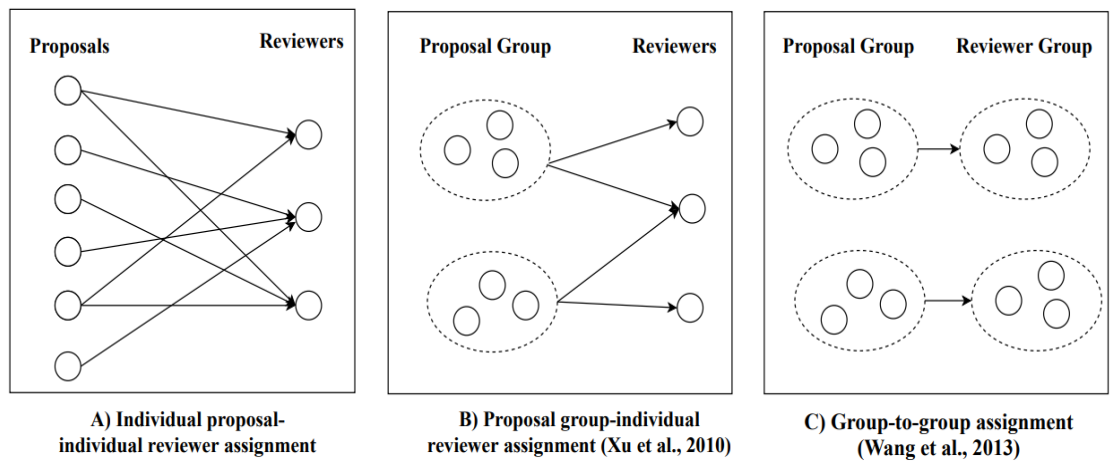

Figure 11: Grouping strategies for assignment.

### 7.3.2 Two-Stage Reviewing Strategy

In large-scale conferences, it becomes difficult to identify the appropriate proposal-reviewer pairs and to fight malicious behavior that hinders the quality of the evaluation. Conferences such as AAAI'21-22 and IJCAI'22 have adopted a two-stage reviewing process system to cope with these challenges under time constraints. In this system, after the first evaluations are made, a subset of the proposals is assigned to additional reviewers for a second review. The main reason for dividing the process into two stages is to use the reviewer resources more effectively and improve the evaluation quality. Under this general goal, two-stage reviewing can be applied in different ways to respond to different needs. For example, as implemented at the AAAI'21, only proposals that have received high enough scores in the first phase and have a chance to be accepted can be examined in the second stage. The two-stage reviewing can also aid in the evaluation of "borderline proposals" that are on the edge between approval and rejection (Leyton-Brown, 2022). It also helps to provide meaningful feedback to applicants of proposals rejected in the first phase. In addition, the two-stage reviewing can be used to compensate the reviewers who did not respond or gave minimal response at the first stage, could not evaluate due to problems in their personal lives, and refused to assess the proposal due to COIs with the proposals assigned to them.

In two-stage reviewing, the proposals that require additional examination are not known in advance. Under severe time constraints, it is difficult to determine and assign the reviewers to the proposals in the second stage. Jecmen et al. (2021) presented a simple "random splitting" strategy that equally randomly selects a subset of reviewers to assign in the second stage. Given a set $P$ consisting of $m$ proposals and set $R$ consisting of $n$ reviewers, a subset of proposals $P_2 \subseteq P$ requires extra evaluation. They selected a subset $R_2 \subseteq R$ of reviewers for the second stage without examined $P_2$, leaving reviewers $R_1 = R \backslash R_2$ to be assigned to proposals in the first stage. They formulated two-stage RAP to maximize the sum of similarity scores of the reviewer assignment across both stages. They used the mean similarity for the optimal assignment to interpret the results easily. Equation 7-17 presents how the optimal assignment mean similarity in the second stage under the constraints (equations. 7-18, 7-19, 7-20) is calculated.

$Q(R_2, P_2) =$

$$\frac{1}{d^{(1)}n + d^{(2)}\beta n}\left[ max_{x \in M(R \backslash R_2, P; w; d^{(1)})} \sum_{i \epsilon P, j \epsilon R \backslash R_2} x_{ij}s_{ij} + max_{y \in M(R_2, P_2; w; d^{(2)})} \sum_{i \epsilon P_2, j \epsilon R_2} y_{ij}s_{ij} \right] \qquad (7\text{-}17)$$

$$s.t. \quad \sum_{i \epsilon P} x_{ij} + y_{ij} \leq w \quad \forall j \in R \qquad (7\text{-}18)$$

$$w|R_2| \geq d^{(2)}\beta n \qquad (7\text{-}19)$$

$$w(m - |R_2|) \geq d^{(1)}\beta n \qquad (7\text{-}20)$$

where $x_{ij}$ is a matrix in which $x_{ij} = 1$ if reviewer $j$ is assigned to proposal $i$ in the first stage and $x_{ij} = 0$ otherwise. $y_{ij}$ is the matrix used for the same purpose in the second stage. $s_{ij}$ is the similarity score between reviewers and proposals. $d^{(1)}$ represents the proposal demand of the first stage and $d^{(2)}$ represents the proposal demand of the second stage. $w$ denotes the maximum reviewer workload. $\beta$ is the fixed value for the proposal fraction and $\beta \in \left\{ \frac{1}{n}, ..., \frac{n}{n} \right\}$.

Leyton-Brown et al. (2022) introduced a new two-stage review system to use available reviewer resources more efficiently by assigning fewer reviewers to proposals with a low probability of acceptance. Two-stage RAP was formulated as a MIP problem that maximizes the overall weighted matching score, which is subject to proposal and reviewer capacity constraints and avoids COIs. Also, this study considered additional soft desiderates such as reviewers' seniority per the proposal, co-authorship distance between reviewers of a proposal, geographic diversity among reviewers assigned to a proposal, and two-cycle free assignment. To guarantee that the resulting MIP formulation may be solved within an acceptable time and memory, they used row/column generation and solved it separately for each stage.

## 8. Concluding Remarks and Future Studies

Peer review is the most widely accepted procedure for evaluating and determining the quality of proposals submitted to academic conferences, journals, or grant organizations. The peer review procedure includes several steps. The assignment of a competent reviewer to the proposal, namely the RAP, is the first and the most critical task for an accurate and fair review process. In the routine peer-review procedure, the conference chair, journal editors, or grant managers must look over all proposals to find suitable reviewers and manually match reviewers with proposals, keeping in mind certain restrictions. As the number of proposals keeps increasing, the old ways of manually assigning reviewers for each proposal do not work anymore because they are time-consuming and prone to mistakes. Due to the evident limits of manual assignment, it is necessary to design automatic reviewer assignment mechanisms. Hence, automatic reviewer assignment has attracted significant academic interest as a research and practical problem.

Over the past 30 years, scholars from different fields have developed reviewer assignment approaches to address the RAP. This study conducted a systematic survey of the proposed automatic procedures in the literature. Numerous methodologies were categorized and discussed according to their advantages and disadvantages. As this review suggests, assigning reviewers to proposals is a field where much progress has been made, and the number of studies on this subject has increased in recent years. However, there are still gaps for improvement in reviewer assignment processing. In this section, the issues on which future studies can focus are listed under the following four headings.

## 8.1 Future Directions for Reviewer and Proposal Profiling

One of the objectives of this literature review is to examine the existing research on the variables that comprise the reviewer and proposal profile and how they are determined. Although many different variables have been considered for reviewer and proposal representation, there are still gaps to consider.

- The majority of the RAP literature assumes that the reviewer candidate database is pre-existing, and they do not describe the database construction process. Further, the process of creating the reviewer candidate databases should be examined. It is also assumed that reviewers in existing databases are always eligible for proposals and available to review. Due to heavy workloads or tight schedules, reviewers who are considered ready to evaluate may not accept this task. Moreover, this approach may result in a shortage of expertise in particular research fields, even when the qualifications of potential reviewers can be ensured. Therefore, the pool of reviewer candidates should be kept as large as possible. It is believed that using information retrieval techniques to recruit new reviewers from external sources would be beneficial for both the expansion of the pool and access to objective and up-to-date information about the reviewers. This deserves in-depth research as well.

- Measuring reviewers' historical evaluation performance and taking it into account in the reviewer assignment task can improve the accuracy and quality of the assignment. To consider the historical evaluation performance of reviewers in the reviewer assignment task, an empirical study can be conducted to determine metrics for assessing the evaluation ability of reviewers, such as evaluation accuracy, quality, and so on.

- In order to model the proposal profile, a few studies have considered some applicant aspects, such as the applicant's institution, area of expertise, and co-authorship information. Future studies should find and use more applicant features, such as educational background, work experience, research projects, and patents representing the applicant's expertise. This kind of information can aid in relevance determination and reviewer connectivity analysis.

- Establishing comprehensive and reliable reviewer and proposal databases is essential to ensure accurate proposal-reviewer matching. It is common in this area to use static databases of manually collected and not updated information. There are also dynamic databases, such as OpenReview and DBLP, that contain structured or unstructured data from multiple sources. However, more efforts are needed to improve the accuracy and completeness of dynamic databases.

## 8.2 Future Directions for Similarity Score Computation

The primary goal of stage 2 is to calculate a similarity score for every possible proposal-reviewer pair based on the profile of candidate reviewers and proposals. In this review paper, the existing studies for calculating similarity scores were summarized into three categories according to the usage of different types of information. While many similarity calculation systems rely on manual approaches, it has been demonstrated that the state-of-the-art tools from NLP technologies significantly automate and enhance similarity score calculation. Besides text-based approaches, a lot of research has proposed network-based models based on implicit relationships. Analyzing such relationships has been used to strengthen the reviewer recommendation and to avoid any potential COIs between proposals and reviewers. There are many efforts to calculate similarity scores between the proposals and the reviewers in the current studies, but further work should be done.

- Neural network-based language models produce outstanding results in a variety of NLP downstream tasks, including document categorization, named entity recognition, and machine translation (Romanov and Khusainova, 2019; Kalyan et al., 2021; Kapočiūtė-Dzikienė et al., 2021). Few studies on the application of neural network-based models to the calculation of similarity scores have shown that they are more effective than other semantic-based methods (Ogunleye et al., 2018; Duan et al., 2019; Yong et al., 2019). However, there is no consensus on the benefits of neural network-based language models in the RAP field. Therefore, more research is needed to analyze the effectiveness of using next-generation embedding models for the RAP. It is thought that particularly transformer-based pre-trained models, such as XLNet (Yang et al., 2019), RoBERTa (Liu et al., 2019), and T5 (Raffel et al., 2020), offer potent feature extraction capabilities that can aid in enhancing the similarity score accuracy between proposals and reviewers.

- Pre-trained language models can be directly used to assign reviewers in any field based on a huge amount of data about proposals and reviewers. However, the quality of the results produced by such learning models depends on the availability of an extensive collection of domain-specific training data. Therefore, researchers and those who arrange peer reviews must collaborate to create and distribute large training datasets.

- Some researchers have proposed a set of similarity score calculation methods in which they use the word and semantic information together. The results were satisfactory when compared to the similarity score calculated based on the word or semantic information alone. In the future, additional hybrid methods must be developed for calculating similarity scores. Also, to improve similarity score computation and the speed and fairness of reviewer assignment, future studies may examine integrating text-based, network-based, and manual-based information.

- Prior studies generally proposed automatically detecting COIs using weighted connections such as co-author and citation networks. However, other social networks, such as academic relationships, advisee-advisor relationships, competitor relationships, and project collaborators, may influence the reviewer's assignment. For instance, COIs may exist between two researchers with the same adviser who have never published together. These potential COIs of applicants and reviewers are self-declared and can be evaluated subjectively, but they could affect the quality of a peer review process. To automatically detect this kind of COIs, supervised learning models can be used to learn the COIs scores between users. But, the absence of high-quality ground-truth datasets severely hinders the advancement of supervised COIs learning. Future research should concentrate on creating ground-truth datasets.

- For accurate COIs detection, it is required to build interactive systems such as PISTIS (Wu et al., 2018), which can automatically extract certain forms of latent COIs from open sources and contain declaration modules for subjective COIs categories.

- In the case of existing COIs between the reviewer and the proposal, it is generally believed that the reviewer should not be assigned to the proposal, regardless of the kind or degree of the COIs. However, this universal approach may result in a lack of qualified reviewers for evaluation, particularly in areas where qualified reviewers are scarce. Instead of this one-size-fits-all approach, flexible COIs avoidance strategies can be developed by considering diverse types and degrees of COIs. Current studies in the RAP literature have focused on identifying and avoiding potential COIs, but how to measure the strength of relationships that can lead to COIs was seldom concerned. Future studies may focus on developing measurement metrics for the degree of relationships. For example, let's say that a reviewer and an applicant of the proposal are two separate nodes. The distance of two nodes can be utilized to estimate the degree of COIs; that is, the closer relationships a reviewer and an applicant share, the more COIs exist between them.

- It is essential to acquire and consolidate social networks from a variety of sources, such as advisee-advisor and affiliation/colleague networks. Consequently, different data sources reflecting various types of relationships must be collected, and the resulting massive dataset must be examined.

## 8.3 Future Directions for Optimizing Assignment

This systematic review demonstrated there are notable studies focused on formulating mathematical models for the RAP to optimize different objective functions under different constraints. In addition, many solution algorithms have been proposed for this problem. But there are still some critical points to consider in the future.

- While the objective function of maximizing overall similarity is quite common in the proposed mathematical model for the RAP, it can lead to injustice for some proposals. This approach can lead to matches involving proposals being assigned to a particular reviewer who lacks expertise in the topical areas of some proposals. To be fair, it is important to ensure that each proposal is assigned to a group of reviewers with the minimum acceptable level of expertise. Recent studies (Kobren et al., 2019; Stelmakh et al., 2019) have attempted to overcome these problems by imposing strict constraints on the minimum proximity of valid proposal-reviewer matches or optimizing the sum of the proximities of a proposal in the worst case. However, it is difficult to find the ideal solution due to the complexity of constraints. Future research should focus on enhancing the computing efficiency of the proposed approach.

- Re-optimization strategies are also deserving of attention. If unknown COIs between the reviewer and proposal are discovered, or the reviewer is severely unqualified for the review, new reviewers must be identified, and a new decision-making procedure must be implemented. In a time-constrained context, it would be advantageous to develop re-optimization approaches rather than repeat the entire procedure.

- It is believed that receiving feedback from reviewers will help to improve reviewers' satisfaction and result in more accurate reviewer-proposal matches. Another future research direction is how to dynamically optimize assignments, including reviewers' feedback on assignment results.

## 8.4 Future Directions for Other Concerns

- This survey showed that the majority of the RAP studies focus on one or two stages, and techniques for resolving different phases are rarely coordinated. Nevertheless, there are inherent relationships between stages. For instance, the assignment outcomes of algorithms in stage (3) depend heavily on the accuracy of similarity score calculations in stage (2). In the future, researchers should develop new methods that consider all three stages instead of just trying to get the best results at each level.

- It will be interesting to see whether the issues of similarity measurement and reviewer assignment can be addressed collaboratively, possibly including feedback from previous issues of the conference or journal or earlier iterations of the grant program.

- First, efforts should be made to establish standardized reviewer assignment criteria and procedures for each area to ensure the accuracy and fairness of the assignment.

- In most studies, real data obtained by random or human assignment was used to evaluate and test proposed methods. These datasets are not accessible publicly because they include sensitive information. Consequently, a direct comparison cannot be made. The absence of a standard benchmark for the RAP impedes comparing the performance of various approaches and, thus, the advancement of research. Therefore, a collection of well-designed benchmark datasets should be established.

This literature review gives readers a comprehensive overview of the RAP as well as cutting-edge RAP-related strategies. In addition, this paper presents a basis for all the phases of the RAP and the classification of the latest studies addressing the RAP. Recent research progress in this field was examined systematically, and open research areas were highlighted. Based on this paper, scholars can consider the literature gaps and recognize potential topics for future research. Furthermore, grant organizations, scientific journals, and conferences can better understand how the reviewer assignment process can be managed more efficiently. Our greatest desire is that this study will be an introductory document for researchers considering work on the RAP.

## Appendix A:  Available Datasets

| Author(s) | Case Implementation | Available Dataset | Dataset Content |
|-----------|---------------------|-------------------|-----------------|
| Karimzadehgan et al. (2008) | Conference Reviewer Assignment: ACM, SIGIR | http://sifaka.cs.uiuc.edu/ir/data/review.html | 73 papers, 189 reviewers |
| Karimzadehgan & Zhai (2009) | Conference Reviewer Assignment: SIGIR | http://timan.cs.uiuc.edu/data/review.html | 73 papers, 30 reviewers |
| Tang et al. (2010) | Conference Reviewer Assignment: KDD'08, '09 and ICDM'09 | http://www.arnetminer.org/expertisematching | 338 papers, 354 reviewers |
| Daud et al. (2010) | Journal Reviewer Assignment: DBLP | http://www.informatik.uni-trier.de/~ley/db/ | 90124 papers, 112317 authors |
| Tang et al. (2012) | Conference Reviewer Assignment: | http://www.arnetminer.org/expertisematching | 338 papers, 354 reviewers |

| Author(s) | Case Implementation | Available Dataset | Dataset Content |
|---|---|---|---|
| | KDD'08,'09 and ICDM'09 | | |
| Karimzadehgan & Zhai (2012) | Conference Reviewer Assignment: ACM | http://timan.cs.uiuc.edu/data/review.html http://www.acm.org/dl. | 73 papers, 189 reviewers |
| Kou et al. (2015a, 2015b) | Conference Reviewer Assignment: SIGKDD'08, '09, ICDM'09 SDM'08, '09 CIKM'08,'09 | http://degroup.cis.umac.mo/reviewerassignment/ | First dataset: 545 papers, 203 reviewers; Second dataset: 648 papers, 154 reviewers; Third dataset: 617 papers, 105 reviewers; Fourth dataset: 513 papers, 90 reviewers |
| Nguyen et al. (2018) | Conference Reviewer Assignment: CCIA'14, '15, '16 | aminer.org, 3tdx.cat, researchgate.net, dblp.uni-trier.de | 106 papers, 96 reviewers |
| Duan et al. (2019) | Journal Reviewer Assignment: Arxiv | http://sifaka.cs.uiuc.edu/ir/data/review.html https://www.aminer.cn/expertisematching ftp://3lib.org//oai_dc/arxiv | 685 papers, 1885 reviewers |
| Mirzaei et al. (2019) | Conference Reviewer Assignment: SIGIR, PubMed | http://timan.cs.uiuc.edu/data/review.html, http://isl.ce.sharif.edu/pubmed/dataset/, http://www.ncbi.nlm.nih.gov/pubmed | SIGIR- 73 papers, 189 reviewers PubMed- 231 papers |
| Kobren et al. (2019) | Conference Reviewer Assignment: MIDL, CVPR | https://github.com/iesl/fair-matching | MIDL: 118 papers, 117 reviewers CVPR: 2623 papers, 1737 reviewers |
| Kalmukov (2020) | Conference Reviewer Assignment: CompSysTech | http://www.compsystech.org/ | 1484 papers |
| Jecmen et al. (2020) | Conference Reviewer Assignment: ICLR'18 | https://github.com/theryanl/mitigating_manipulation_via_randomized_reviewer_assignment | 911 papers, 2435 reviewers |
| Kreutz & Schenkel (2020) | Conference Reviewer Assignment: MOL'17, BTW'17, ECIR'17 | https://doi.org/10.5281/zenodo.4071874 | _ |
| Pradhan et al. (2020) | Conference Reviewer Assignment: ICBIM'16 | https://easychair.org/my/conference.cgi?welcome=1;conf=icbim2016 | 59 papers, 40 reviewers |
| Xu et al. (2020) | Conference Reviewer Assignment: ICLR'18 | https://github.com/xycforgithub/StrategyProof_Conference_Review | 911 papers, 2435 reviewers |
| Tan et al. (2021) | Journal Reviewer Assignment: Arxiv, GitHub | https://github.com/aitsc/WSIM/tree/main/datasets | 100 papers, 400 reviewers |
| Hoang et al. (2021) | Journal Reviewer Assignment: DBLP | https://dblp.org/xml/, https://github.com/Lucaweihs/impact-prediction#getting-data | 479 papers, 1000 reviewers |
| Patil (2021) | Conference Reviewer Assignment: NIPS'19 | https://www.kaggle.com/abolihpatil/nips2019-paper-for-reviewer-assignment-ahp-pnm | 150 papers |

## Appendix B:  Text-Based/Content-Based Approaches

| Author(s) | Case Implementation | Reviewer Profile | Proposal Profile | Methods | Performance Metric Used | Comparison |
|---|---|---|---|---|---|---|
| Dumais & Nielsen (1992) | Conference Reviewer Assignment: Hypertext'91 | Abstracts provided by reviewers | Title, abstract of proposals | LSI | Average relevance, Precision | Comparison with reviewers' relevance judgments |

| Author(s) | Case Implementation | Reviewer Profile | Proposal Profile | Methods | Performance Metric Used | Comparison |
|---|---|---|---|---|---|---|
| Yarowsky & Florian (1990) | Conference Reviewer Assignment: ACL'99 | Full text, abstract, title, keywords of reviewers' previously published papers, the bibliographic information of reviewers (co-authorship, citations) | Full-text, abstracts, title, keywords of proposals; bibliographic information of the author | VSM, Naive Bayes Classifier | Accuracy, Average position, One-of-best-2 | 1. Recommended model comparison with manual assignment, 2. Comparison of assignment made using text, bibliography, text + bibliography information, 3. The case where the full texts of reviewers' publications are used with equal weight, different weights, and manual assignment comparison, 4. Assignment using only keywords vs. assignment using keyword, title, abstract comparison |
| Basu et al. (2001) | Conference Reviewer Assignment: AAAI-98 | Full-text of reviewers' previously published papers, reviewers' names, affiliations, preferences | Abstract of proposals | TF-IDF, KNN, Extended Direct Bayes | Recall | 1. Recommended algorithms comparison with each other 2. Comparison with random assignment and collaborative filtering |
| Hettich & Pazzani (2006) | Project Proposal Reviewer Assignment: NSF | Full-text of reviewers' previously published papers | Full-text of proposals | TF-IDF, Hill climbing algorithm | Sum of residual term weight | Comparison among alternative reviewer set and proposal set |
| Ferilli et al. (2006) | Conference Reviewer Assignment: IEA/AIE | Abstract and title of reviewers' previously published papers | Title, abstract of proposals | LSI | Accuracy | NA |
| Biswas & Hasan (2007) | Journal Reviewer Assignment: Simulated data | Full-text of reviewers' previously published papers, domain ontology of reviewers | Full-text of proposals | VSM, KEA | Relevance | Comparison among free-text, KEA, ontology-driven topic inference |
| Zhao et al. (2018) | Journal Reviewer Assignment: Simulated data | The research interest of reviewers | Keywords of proposals | Word Mover's Distance, Constructive Covering Algorithm, LDA–KNN, LDA–CCA, LSI–KNN, LSI–CCA, | Precision, Accuracy, Recall | 1. Comparison of the synthetic dataset and public dataset 2. Comparison of LDA–KNN, LDA–CCA, LSI–KNN, LSI–CCA, WMD–KNN, LDA–GaussianNB, LDA–SVM, LSI–GaussianNB |

| Author(s) | Case Implementation | Reviewer Profile | Proposal Profile | Methods | Performance Metric Used | Comparison |
|---|---|---|---|---|---|---|
| | | | | WMD–KNN, LDA–GaussianNB, LDA–SVM, LSI–GaussianNB | | |
| Mimno & Mccallum (2007) | Conference Reviewer Assignment: NIPS'06 | Full-text of reviewers' previously published papers | Full-text of proposals | APT | Accuracy, Precision | 1. Recommended model comparison with manual assignment, 2. Comparison among Dirichlet smoothed language model, APT, and ATM |
| Karimzad ehgan et al. (2008) | Conference Reviewer Assignment: ACM, SIGIR | Abstract of reviewers' previously published papers | Abstract, full-text of proposals | PLSA, k-means | Confidence, Coverage, Average Confidence, F-score | 1. Reviewer aspect modeling (maximal marginal relevance vs. PLSA) comparison with Paper aspect modeling (mutual information-based clustering vs. simple discourse analysis), 2. Comparison of abstract and full-text usage |
| Tang et al. (2008) | Conference Reviewer Assignment: Open databases of publications | Full-text of reviewers' previously published papers, the bibliographic information of reviewers (co-authorship, citations) | NA | ACT | Mean average precision | ACR with three strategies are compared with SLM, LDA, ATM |
| Conry et al. (2009) | Conference Reviewer Assignment: ICDM'07 | Number of co-authored publications, research area, preferences of reviewers, and abstract of reviewers' previously published papers | Abstract and research area of proposals | VSM | Evaluation ratings | Comparison of the proposed model and Taylor's (2008) model |
| Tang et al. (2010) | Conference Reviewer Assignment: KDD'08, '09 and ICDM'09 | Reviewers' previously published papers, co-authorship, affiliation | NA | Language Model, LDA | Precision | Comparison of the proposed algorithm and greedy algorithm |
| Daud et al. (2010) | Journal Reviewer Assignment: DBLP | Reviewers' previously published papers | NA | Temporal-Expert-Topic Model | Entropy of topics | Comparison of the proposed Temporal-Expert-Topic approach with two baseline |

| Author(s) | Case Implementation | Reviewer Profile | Proposal Profile | Methods | Performance Metric Used | Comparison |
|---|---|---|---|---|---|---|
| | | | | | | approaches, Temporal-Author-Topic and Author-Conference-Topic |
| Andrade-Navarro et al. (2012) | Journal Reviewer Assignment: MEDLINE | Abstract and reference of reviewers' previously published papers, co-authorship | Title and abstract of proposals | Fuzzy binary relations | Average score of similarity, Average papers per reviewer profile | 1. Comparison of Peer2ref and manual reviewer selection results 2. Peer2ref, eTBlast, and Jane web tools comparison |
| Mai et al. (2012) | Project Proposal Reviewer Assignment: NSFC | Research area code, evaluation experience, project application experience | Title, abstract, and keywords of proposals | Probabilistic language models | Bidding | Comparison of two proposal sets: popular and unpopular topic |
| Xue et al. (2012) | Conference Reviewer Assignment: KDD'08, '09 and ICDM'09 | Full-text of reviewers' previously published papers, reviewers' names, affiliations | Discipline ontology, keywords of proposals | Interval fuzzy ontology, Semantic matching | Matching degree, Load variance | Topic number comparison |
| Charlin et al. (2012) | Conference Reviewer Assignment: NIPS'09, 10 | Keywords provided by reviewers, citations, and experience of reviewers | Keywords and citations of proposals | Language model, Linear Regression, Bayesian probabilistic matrix factorization | RMSE | Comparison of the language model, linear regression, Bayesian probabilistic matrix factorization |
| Li & Watanabe (2013) | Research Project Proposal Reviewer Assignment: NSFC | Number of publications, quality and freshness of publications, journal ranking, references of reviewers | References of proposals | APT | NA | NA |
| Long et al. (2013) | Conference Reviewer Assignment: KDD'06,10 | NA | NA | TF-IDF | Fairness of assignment | Assignment comparison by different types of conflicts of interest |
| Charlin & Zemel (2013) | Conference Reviewer Assignment: NIPS'10, ICML'12 | Full-text of reviewers' previously published papers, expertise score provided by reviewers | Full-text of proposals | Language model, LDA, Linear Regression, Bayesian probabilistic matrix factorization | Relevance score, RMSE | 1. Language model vs. LDA comparison, 2. Linear Regression vs. Bayesian probabilistic matrix factorization comparison |

| Author(s) | Case Implementation | Reviewer Profile | Proposal Profile | Methods | Performance Metric Used | Comparison |
|---|---|---|---|---|---|---|
| Caldera et al. (2014) | Conference Reviewer Assignment: Eurographics' 14 | Full-text of reviewers' previously published papers | Full-text of proposals | TF/IDF | Preference of reviewers | NA |
| Liu et al. (2014) | Conference Reviewer Assignment: CIKM'08, NIPS'06 | Expertise topic, co-authorship of reviewers, Full-text of reviewers' previously published papers | Full text, co-authorship relationship of proposals | LDA | Precision, Matching degree, Authority, Diversity | 1. Comparison of NIPS and CIKM datasets 2. Comparison of RWR, topic similarity, text similarity, RWR with a sparsity constraint |
| Protasiewicz (2014) | Journal Reviewer Assignment: Open databases of publications | Full-text, abstracts, title, keywords of reviewers' previously published papers; name-surname, affiliation, research area, co-authorship of reviewers | Full-text of proposals | Multinomial Naïve Bayes, SVM, Rapid Automatic Keyword Extraction Algorithm | NA | NA |
| Kou et al. (2015a, 2015b) | Conference Reviewer Assignment: SIGKDD'08, '09, ICDM'09 SDM'08, '09 CIKM'08,'09 | Abstract of reviewers' previously published papers | Abstract of proposals | ATM | NA | NA |
| Protasiewicz et al. (2016) | Project Proposal Reviewer Assignment: National Centre for Research and Development of Poland | Full-text, abstracts, title, keywords of reviewers' previously published papers; name-surname, affiliation, expertise area, co-authorship of reviewers | Title, abstract, and full-text of proposals | Naive Bayes, Hierarchical Agglomerative Clustering, | Precision, Accuracy, F-score | 1. Comparison of Polish keyword extraction, keyphrases extraction algorithm, rapid automatic keyword extraction 2. Maximum probability model comparison with multilingual classification |
| Shon et al. (2016) | Project Proposal Reviewer Assignment: Korea NSF | Abstract, title, and keywords of reviewers' previously published papers, Affiliation of reviewers | Title, abstract, and keywords of proposals; keywords provided by applicants | TF-IDF | Execute time | NA |

| Author(s) | Case Implementation | Reviewer Profile | Proposal Profile | Methods | Performance Metric Used | Comparison |
|---|---|---|---|---|---|---|
| Yin et al. (2016) | Journal Reviewer Assignment: DBLP | Full-text of reviewers' previously published papers, h-index of reviewers | Full-text of proposals | Topic Model | Diversity, F-score, Coverage, Authority | Comparison among proposed model, model without taking into consideration the reviewers' factor importance and reviewer aspect model |
| Jin et al. (2017) | Journal Reviewer Assignment: WANFANG, ArnetMiner | NA | Full-text of proposals | ATM, EM | Relevance, Interest trend, Authority | 1. Comparison of ATM, SLM, and VSM, 2. Comparison of two datasets: WANFANG and ArnetMiner |
| Peng et al. (2017) | Journal Reviewer Assignment: Frontiers of Computer Science Journal | Full-text, abstract, title, and time information of reviewers' previously published papers | Title, abstract, and full-text of proposals | LDA, TF-IDF | Matching degree | Comparison of the proposed method and manual assignment |
| Ogunleye et al. (2017) | Conference Reviewer Assignment: NIPS | Full-text of reviewers' previously published papers | Full-text of proposals | Word2Vec, TF-IDF, LSI, LDA | Suitability score | Comparison among Word2Vec, TF-IDF, LSI, LDA |
| Yan et al. (2017) | Conference Reviewer Assignment: ICDM'15, SIGKDD'13,14,15 | Full-text of reviewers' previously published papers | Abstract of authors' previously published papers | LDA | Average matching fitness, Average avoiding effect | Comparison of the different coupling parameter value |
| Jin et al. (2018a) | Journal Reviewer Assignment: WANFANG, ArnetMiner | Abstract of reviewers' previously published papers, co-authorship of reviewers | NA | AST | Perplexity, Kullback–Leibler divergence, Coverage | Comparison of two different types of AST and ATM |
| Li & Hou (2018) | Conference Reviewer Assignment: IEEE INFOCOM'15,16 | Full-text of reviewers' previously published papers | Full-text of proposals | LSI | NA | NA |
| Nguyen et al. (2018) | Conference Reviewer Assignment: CCIA'14, '15, '16 | Research topic, interests, recency, and quality of reviewers | Full-text of proposals | LDA | Quality index, Overall matching | Comparison of the proposed method and random assignment |
| Anjum et al. (2019) | Conference Reviewer Assignment: NIPS, Tier-1 | Abstract of reviewers' previously published papers | Abstract of proposals | Word2Vec, Common Topic Model | Precision | Comparison of APT, Single Doc model, Hierarchical Dirichlet Process, Random Walk with Restart, LDA, Word |

| Author(s) | Case Implementation | Reviewer Profile | Proposal Profile | Methods | Performance Metric Used | Comparison |
|---|---|---|---|---|---|---|
| | | | | | | Mover's Distance, Hidden Topic Model, Doc2Vec, and proposed model |
| Medakene et al. (2019) | Conference Reviewer Assignment: ICA2IT'19 | References of reviewers' previously published papers, pedagogical facet of reviewers, research area | Full-text and references of proposals | LDA | Confidence degree, Topic similarity, Relevance score, MSE, RMSE, MAE, MAPE | Comparison of the situation with and without consideration of the pedagogical profile |
| Duan et al. (2019) | Journal Reviewer Assignment: Synthetic and Arxiv | Title and abstract of reviewers' previously published papers | Title and abstract of proposals | TF-IDF, BERT, CNN, biLSTM | Precision, Recall, Macro-F1 score, Mean Average, Precision, Normalized discounted cumulative gain, Binary preference | 1. Comparison of SPM-RA, LDA, SLM, LDA-LM, TATB, Keyword cosines similarity, BBA, WMD algorithm 2. Comparison of the synthetic dataset and public dataset 3. Comparison between six neural network structures for the sentence pair model: Word2Vec-bilSTM, Word2Vec-CNN, Word2Vec-bilSTM-CNN, bilSTM, CNN, bilSTM-CNN |
| Mirzaei et al. (2019) | Conference Reviewer Assignment: AGM SIGIR, PubMed | Title and abstract of reviewers' previously published papers | NA | PLSA, LSA | Coverage, Average confidence | Comparison among ILP-PLSA, CFLA-PLSA, WCGRA-PLSA, CMARTA-LRA-ILP, CMARTA-LRA |
| Mittal et al. (2019) | Conference Reviewer Assignment: SemEVAL'10 | Full-text of reviewers' previously published papers | Full-text of proposals | Wordnet, Fuzzy Extension Principle Fuzzy Graph Connectivity Measures | Matching degree | Comparison of fuzzy degree centrality, fuzzy betweenness centrality, fuzzy closeness centrality, PageRank, HIT |
| Chughtai et al. (2019) | Journal Reviewer Assignment | Full-text of reviewers' previously published papers | Title, abstract, keywords of proposals | LSA, TF/IDF, Ontology-based knowledge-domain | Entropy, Similarity index | NA |
| Zhang et al. (2019) | Conference Reviewer Assignment: ACM | Title and abstract of reviewers' previously published papers | Title and abstract of proposals | Hiepar | Recall, Accuracy, Normalized discounted cumulative gain | 1. Comparison of the proposed model, LDA, Word2vec, CNN, BERT, Bi-LSTM, and LSTM 2. Comparison of Hiepar with TF/IDF, LSI, LDA, WMD, BBA |

| Author(s) | Case Implementation | Reviewer Profile | Proposal Profile | Methods | Performance Metric Used | Comparison |
|---|---|---|---|---|---|---|
| Kalmukov (2020) | Conference Reviewer Assignment: CmpSysTech | Abstract of reviewers' previously published papers | Abstract of proposals | VSM, TF-IDF | Preferences of reviewers | Comparison of different TF-IDF weighting models |
| Yang et al. (2020) | Conference Reviewer Assignment: WWW, SIGIR, CIKM | Title, keywords, and abstract of reviewers' previously published papers | Abstract and keywords of proposals | LDA | Computational time, Objective function value, Matching degree | NA |
| Pradhan et al. (2020) | Conference Reviewer Assignment: CBIM'16 | Expertise area, co-authorship of reviewers, Full-text of reviewers' previously published papers | Title of proposals | LDA | Quality | Comparison of the proposed method with EasyChair and TPMS systems |
| Tan et al. (2021) | Journal Reviewer Assignment: Arxiv, GitHub | Title and abstract of reviewers' previously published papers | Title and abstract of proposals | Language Model, LDA, Word, and semantic-based iterative model (WSIM) | Accuracy, Precision, Recall, Macro-F1 score, Mean averaged precision, Binary preference, Normalized discounted cumulative gain | Comparison of LDA, SLM, LDA-SLM, Word2Vec, LDA-Branch and bound, time-aware, and topic-based keyword cosine similarity |
| Hoang et al. (2021) | Journal Reviewer Assignment: DBLP | Full text of publications, number of publications, number of citations, h-index of reviewers | Full-text of proposals | LDA, Doc2Vec | Normalized discounted cumulative gain, Discounted cumulative gain | Studies of Protasiewicz (2014), Hoang et al. (2019), and proposed model comparison |
| Patil & Mahalle (2021) | Conference Reviewer Assignment: Interspeech, NIPS'19, AAAI'14 | Name, affiliation, publications, and network information of reviewers | Title, abstract, keywords of proposals | LDA | NA | NA |

## Appendix C:  Optimization-Based Approaches

| Author(s) | Case Implementation | Mathematical Model | Algorithm | Performance Metrics Used | Comparison |
|---|---|---|---|---|---|
| Hartvigsen et al. (1999) | Conference Reviewer Assignment | Maximum weight-capacitated transportation problem | NA | Computational time | NA |
| Guerv´os & Valdivieso (2004) | Conference Reviewer Assignment: PPSN2002 | GAP(ILP) | Greedy algorithm & Evolutionary algorithm | NA | NA |
| Cook et al. (2005) | Project Proposal Reviewer Assignment: Simulated data | Set covering integer programming | Greedy Algorithm | Computational time, Solution quality | 1. Comparison of 4 sample datasets with different sizes, 2. Comparison of set covering integer programming and greedy algorithm |
| Hettich & Pazzani (2006) | Project Proposal Reviewer Assignment: NSF | NA | Hill climbing algorithm | Sum of residual term weight | Comparison of alternative sets of reviewers and alternative approaches |
| Janak et al. (2005) | Research Project Proposal Reviewer Assignment: NSF | GAP(ILP/MIP) | NA (Using CPLEX) | Objective function value | 1. Comparison of recommended model and manual assignment, 2. Comparison of four sample datasets with different sizes |
| Sun et al. (2007) | R&D Project Proposal Reviewer Assignment: NSFC | GAP(ILP) | Simplex algorithm | Expertise level, Workload of reviewers | NA |
| Li et al. (2007) | Project Proposal Reviewer Assignment: Tianjin NSF | GAP(ILP) | Genetic Algorithm | Average fitness value, The best fitness value | Comparison of standard genetic algorithm, genetic algorithm based on mechanism of positive feedback and adaptive mutation by pheromone |
| Goldsmith & Sloan (2007) | Conference Reviewer Assignment | Network flow | NA | NA | NA |
| Li et al. (2008) | Project Proposal Reviewer Assignment: Tianjin NSF | Complicated discrete optimization problem | Ant colony optimization | Average fitness value, The best fitness value | Comparison of ant colony optimization genetic algorithm based on mechanism of positive feedback and adptive mutation by pheromone |
| Karimzadehgan & Zhai (2009) | Journal Reviewer Assignment: Simulated data | GAP(ILP) | ILP algorithms, Greedy algorithm | Average confidence | Comparison of ILP algorithm and greedy algorithm |
| Conry et al. (2009) | Conference Reviewer Assignment: ICDM'07 | GAP(ILP) | NA | Evaluation ratings | Comparison of the proposed model and Taylor's (2008) model |

| Author(s) | Case Implementation | Mathematical Model | Algorithm | Performance Metrics Used | Comparison |
|---|---|---|---|---|---|
| Xu et al. (2010) | Research Project Proposal Reviewer Assignment: NSFC | GAP(ILP) | Greedy randomized adaptive search procedure, Genetic algorithm | Average expertise, Workload of reviewers | NA |
| Huang et al. (2010) | Journal Reviewer Assignment: Simulated data | Minimum cost of flow problem | NA | Approximate stability, Satisfaction | NA |
| Tang et al. (2010) | Conference Reviewer Assignment: KDD'08, KDD'09 and ICDM'09 | Minimum convex cost flow problem | Proposed algorithm | Precision | Comparison of the proposed model and greedy algorithm |
| Tang et al. (2012) | Conference Reviewer Assignment: KDD'08, 09 and ICDM'09 | Minimum convex cost flow problem | Proposed algorithm | Matching degree, Load variance, Expertise variance, CPU time | Comparison of authority balance, basic network flow, greedy baseline |
| Chen et al. (2012) | Project Proposal Reviewer Assignment: Simulated data | GAP(ILP) | Genetic algorithm | Preferences of reviewers, Times of movement for reviewers to change venues, Computation time | Comparison of small sample dataset and large sample dataset |
| Karimzadehgan & Zhai (2012) | Conference Reviewer Assignment: ACM Digital library | GAP(ILP) | Greedy algorithm | Coverage, Confidence, Average Confidence | Comparison of greedy algorithm and branch and cut algorithms |
| Xue et al. (2012) | Conference Reviewer Assignment: KDD'08, 09, ICDM'09 | GAP | Hungarian algorithm | NA | NA |
| Charlin et al. (2012) | Conference Reviewer Assignment: NIPS'08, 10 | GAP(ILP) | ILP algorithm | NA | NA |
| Long et al. (2013) | Conference Reviewer Assignment: KDD'06, 10 | MaxTC-PRA | Greedy algorithm for MaxTC-PRA | Fairness of assignment | Assignment comparison by different types of conflicts of interest |
| Tayal et al. (2014) | R&D Project Proposal Reviewer Assignment: Simulated data | Pairwise-constrained problem | Randomized algorithm, Standard assignment algorithm | Matching degree | Comparison of the proposed method and reviewers' preferences |
| Wang et al. (2013) | Journal Reviewer Assignment: Business school in China | GAP(MIP) | Two-phase stochastic-biased greedy algorithm | Total cost, Computational time | Comparison of CPLEX and two-phase stochastic-biased greedy algorithm |

| Author(s) | Case Implementation | Mathematical Model | Algorithm | Performance Metrics Used | Comparison |
|---|---|---|---|---|---|
| Li & Watanabe (2013) | Research Project Proposal Reviewer Assignment: NSFC | GAP(ILP) | Hungarian algorithm | Time consumption, assignments under constraints | Comparison of the proposed model and Hungarian algorithm |
| Li et al. (2013) | Research Project Proposal Reviewer Assignment: A University | GAP(ILP) | Adaptive parallel genetic algorithm | Average computation time, best optimal adaptive value, Average optimal adaptive value | Comparison of adaptive parallel genetic algorithms with two cores, adaptive parallel genetic algorithms with four cores, a random search algorithm, and a genetic algorithm |
| Charlin & Zemel (2013) | Conference Reviewer Assignment: NIPS'10, ICML'12 | GAP(ILP) | NA | NA | NA |
| Cechlarova et al. (2014) | Scientific Project Proposal Reviewer Assignment: Slovak Research and Development Agency | Network flow | NA | The total workload of reviewers | Comparison of the proposed method and fine-tuned model |
| Daş & Gökçen (2014) | R&D Project Proposal Reviewer Assignment: Grant organization | Fuzzy linear programming | NA | Matching degree | Comparison of various degrees of feasibility |
| Silva et al. (2014) | R&D Project Proposal Reviewer Assignment: NSFC | GAP(ILP) | Hungarian algorithm | Accuracy, Recall, Precision, F-score | Comparison among VSM, ATM, LSI, scale-free network, and integer programming (using TF-IDF) |
| Kolasa & Krol (2011) | Conference Reviewer Assignment: KNS'09, KES'09 | GAP | Ant colony optimization & Genetic algorithm, Genetic algorithm & Tabu search algorithm | Computational time, Eligibility degree, Quality of the solution | 1. Comparison of genetic algorithm, tabu search, ant colony optimization, ant colony optimization & genetic algorithm, and genetic algorithm & tabu search 2. KNS, KES, and synthetic dataset comparison |
| Kou et al. (2015a, 2015b) | Conference Reviewer Assignment: SIGKDD'08, 09, ICDM'08, 09, SDM'08, 09, CIKM'08,'09 | WGRAP | Stage deeping greedy algorithm | Response time, Optimality ratio, Superiority ratio | 1. Branch and bound, brute force search and ILP comparison, 2. Comparison of stage deeping greedy algorithm, greedy algorithm, stage deeping greedy algorithm with stochastic refinement, best reviewer group greedy, ILP, and stable matching |

| Author(s) | Case Implementation | Mathematical Model | Algorithm | Performance Metrics Used | Comparison |
|---|---|---|---|---|---|
| Liu et al. (2016) | R&D Project Proposal Reviewer Assignment: NSFC | Capacitated transportation problem | Simplex algorithm | Balance and rationality, Assignment time, Initialization time, Effect on avoiding the interest conflict | Comparison of manual assignment, decision support systems, and internet-based science information systems |
| Yue et al. (2017) | Scientific Project Proposal Reviewer Assignment: NSFC | NP-C problem | Randomized algorithm | Fitness degree, Research intensity, Academic association | NA |
| Ogunleye et al. (2017) | Conference Reviewer Assignment: NIPS | GAP(ILP) | NA | NA | NA |
| Jin et al. (2017) | Journal Reviewer Assignment: WANFANG, ArnetMiner | GAP(ILP) | ILP algorithm | NA | NA |
| Jin et al. (2018b) | Journal Reviewer Assignment: WANFANG, ArnetMiner | GAP(ILP) | ILP algorithm | NA | NA |
| Li & Hou (2018) | Conference Reviewer Assignment: IEEE INFOCOM'15,16 | GAP(ILP) | ILP Solver | NA | NA |
| Lian et al. (2018) | Conference Reviewer Assignment: PrefLib | GAP(MIP) | Network flow algorithm | Mean utility, Hoover index, Gini index | 1. Comparison of Harmonic OWA, Linear OWA, and Geometric OWA<br>2. Comparison of decreasing OWA vectors and increasing OWA vectors<br>3.Comparison of utilitarian and egalitarian |
| Guo et al. (2018) | Conference Reviewer Assignment: SIGIR'07, NIPS'06 | k-loop free assignment | k-loop free assignment algorithm, Integer programming, Greedy algorithm | Total suitability score | 1. Comparison of two datasets: NIPS'06 and SIGIR'07<br>2. Different loop length comparison<br>3. Comparison of assignment with k-loop free constraint and assignment ignoring k-loop free constraint |
| Yeşilçimen & Yıldırım (2019) | R&D Project Proposal Reviewer Assignment: Random sample | GAP(MIP) | OBH algorithm | Computation time, Upper bounds | 1. Comparison of the mathematical model of Cook et al. (2005) and the model presented in this study<br>2. Comparison of linear programming and semi- |

| Author(s) | Case Implementation | Mathematical Model | Algorithm | Performance Metrics Used | Comparison |
|---|---|---|---|---|---|
| | | | | | definite programming relaxation |
| Mirzaei et al. (2019) | Conference Reviewer Assignment: AGM SIGIR, PubMed | Unconstrained Multi-Aspect Review-Team Assignment using Latent Research Areas (UMARTA-LRA), Constrained Multi-Aspect Review-Team Assignment using Latent Research Areas (CMARTA-LRA) | Stage deepening Greedy algorithm, Greedy forward-selection, Stage deepening greedy algorithm-stochastic refinement | Coverage, Average confidence | 1. Comparison of UMARTA-LRA and CFLA 2. Comparison of CMARTA-LRA, CFLA, and ILP 3. Comparison of UMARTA-KL and UMARTA-LRA 4. Comparison of ILP-PLSA, CFLA-PLSA, WCGRA-PLSA, CMARTA-LRA-ILP, CMARTA-LRA |
| Stelmakh et al. (2019) | Conference Reviewer Assignment: CVPR'17, 18, MIDL | GAP(ILP) | PeerReview4All algorithm | Fairness, Sum of similarities, Mean sum of similarities, Computation Time | 1. Comparison of PeerReview4All and the TPMS (Charlin and Zemel, 2013), ILPR (Garg et al., 2010), hard algorithms 2. Comparison of PeerReview4All and the TPMS (Charlin and Zemel, 2013), FAIRIR (Kobren et al., 2019), FAIRFLOW (Kobren et al., 2019) |
| Kobren et al. (2019) | Conference Reviewer Assignment: MIDL, CVPR'17 | GAP(ILP) | FAIRIR (FAIR matching via Iterative Relaxation) FAIRFLOW (min-cost-flow-based heuristic for constructing fair matchings) | Computation time, Paper score, Number of papers assigned to a reviewer, Objective function score, | Comparison of TPMS, FAIRIR, FAIRFLOW, and PR4A algorithms |
| Yang et al. (2020) | Conference Reviewer Assignment: WWW, SIGIR, and CIKM | GAP(ILP) | MBBPSO | Computation time, Fitness function value, Matching degree | Comparison of PSO, BBPSO, MBBPSO, and genetic algorithm |
| Jecmen et al. (2020) | Conference Reviewer Assignment: ICLR'18 | Pairwise-constrained problem | Randomized algorithm | Similarity sum, Assignment probability | 1. Comparison of the pairwise-constrained problem and partition-constrained problem 2. Randomized algorithm vs. standard assignment algorithm comparison |

| Author(s) | Case Implementation | Mathematical Model | Algorithm | Performance Metrics Used | Comparison |
|---|---|---|---|---|---|
| | | | | | 3. Comparison of four sample datasets with different sizes |
| Xu et al. (2020) | Conference Reviewer Assignment: ICLR'17, 18 | Strategyproof-via-partitioning assignment | Divide-and-Rank assignment algorithm | NA | NA |
| Pradhan et al. (2020) | Conference Reviewer Assignment: ICBIM'16 | Equilibrium multi-job assignment | Weighted-matrix factorization-based Greedy Algorithm | Quality | Comparison of the EasyChair, TPMS, and proposed method |
| Kat (2021) | Research Project Proposal Reviewer Assignment: TÜBİTAK | GAP(ILP) | NA | Sum of scores, Mean of rating levels, Ratios of standard deviations of rating percentages to the mean | Comparison of the proposed model and two integer programming models with different objective functions |
| Boehmer et al. (2021) | Conference Reviewer Assignment: ICLR'18 | Weighted z-cycle-free peer reviewing | Greedy algorithm | Fraction of agents | Comparison of ILP and proposed heuristics |
| Jecmen et al. (2021) | Conference Reviewer Assignment: ICLR'18, SIGIR'07, PrefLib3, Bid1 | Two-stage paper assignment problem | Random split | Oracle optimal solution's similarity, Average assignment value | 1. Comparison of different datasets 2. Comparison of the different assignment load parameter |
| Payan & Zick (2022) | Conference Reviewer Assignment: MIDL, CVPR, CVPR'18 | Fair allocation problem | Greedy reviewer round-robin algorithm | Utilitarian social welfare, Nash social welfare, number of EF1 Violation, Minimum paper score, Run time, Sum of the envy across all paper pairs, Gini coefficient of all paper scores | Comparison of GRRR, FairIR, and FairFlow algorithms (Kobren et al., 2019), the TMPS (Charlin and Zemel, 2013), and PeerReview4All (Stelmakh et al., 2019) |
| Dhull et al. (2022) | Conference Reviewer Assignment: ICLR'18 | Strategyproof-via-partitioning assignment | Random partition, Coloring algorithm, Cycle-braking algorithm, multi-partition algorithm | Total similarity, Review score | Comparison of random partitioning, cycle-breaking algorithm, coloring algorithm, multi-partition algorithm, and optimal non-strategyproof assignment |
| Leyton-Brown et al. (2022) | Conference Reviewer | GAP(MIP) | NA | Confidence of reviewer, Percent | 1. Comparison of proposed algorithm and |

| Author(s) | Case Implementation | Mathematical Model | Algorithm | Performance Metrics Used | Comparison |
|---|---|---|---|---|---|
|  | Assignment: AAAI'21 |  |  | reduction of soft constraint violations, Wall time, Number of iterations, Number of papers | matching algorithm of AAAI'20 2. Comparison of the proposed approach and proposed approach without soft constraints and single-phase version of AAAI'21 3. Running time vs. conference size comparison |